\documentclass[acmtog, nonacm]{acmart}

\usepackage{makecell}
\usepackage[most]{tcolorbox}
\usepackage{rotating}
\usepackage{longtable}
\usepackage{booktabs}
\usepackage[table]{xcolor}
\usepackage{color}
\usepackage{graphicx}
\usepackage{caption}
\usepackage{caption}
\usepackage{wrapfig}
\usepackage{multirow}
\usepackage{multicol}
\usepackage{enumitem}
\usepackage{tabularx}
\usepackage{balance}
\usepackage{fvextra}

\DefineVerbatimEnvironment{PromptVerbatim}{Verbatim}{
    breaklines=true,
    breakanywhere=true,
    fontsize=\scriptsize,
    baselinestretch=0.86,
    xleftmargin=0pt,
    xrightmargin=0pt
}
\definecolor{title_purple}{rgb}{0.65,0.1,0.65}

\definecolor{green}{rgb}{0, 0.5, 0}
\definecolor{orange}{rgb}{0.8, 0.6, 0.2}
\definecolor{red}{rgb}{1.0, 0.0, 0.0}
\definecolor{teal}{rgb}{0.0, 0.4, 0.4}
\definecolor{purple}{rgb}{0.65,0.0,0.65}
\definecolor{saffron}{rgb}{0.95,0.75,0.2}
\definecolor{turquoise}{rgb}{0.0,0.5,0.5}
\definecolor{black}{rgb}{0.0, 0.0, 0.0}
\definecolor{gray}{rgb}{0.5, 0.5, 0.5}

\newcommand{\paper}{Sound Sparks Motion}

\definecolor{darkpink}{rgb}{0.561, 0.282, 0.427}
\definecolor{darkturquoise}{rgb}{0., 0.81, 0.822}

\makeatletter
\DeclareRobustCommand\onedot{\futurelet\@let@token\@onedot}

\makeatother

\makeatletter
\def\blfootnote{\xdef\@thefnmark{}\@footnotetext}
\makeatother

\usepackage{color}
\definecolor{green}{rgb}{0, 0.5, 0}
\definecolor{deepred}{rgb}{0.9, 0.15, 0.15}
\definecolor{orange}{rgb}{0.8, 0.6, 0.2}
\definecolor{red}{rgb}{1.0, 0.0, 0.0}
\definecolor{teal}{rgb}{0.0, 0.4, 0.4}
\definecolor{purple}{rgb}{0.65,0,0.65}
\definecolor{saffron}{rgb}{0.75,0.05,0.05}
\definecolor{turquoise}{rgb}{0.0,0.30,0.15}
\definecolor{black}{rgb}{0.0, 0.0, 0.0}
\definecolor{gray}{rgb}{0.5, 0.5, 0.5}

\renewcommand\footnotetextcopyrightpermission[1]{}
\title{Sound Sparks Motion: Audio and Text Tuning for Video Editing}

\author{AmirHossein Razlighi}
\affiliation{
  \institution{University of Cyprus}
  \country{Cyprus}
}
\email{naghirazlighi.amirhossein@ucy.ac.cy}

\author{Aryan Mikaeili}
\affiliation{
  \institution{Simon Fraser University}
  \country{Canada}
}
\email{aryan_mikaeili@sfu.ca}

\author{Ali Mahdavi-Amiri}
\affiliation{
  \institution{Simon Fraser University}
  \country{Canada}
}
\email{amahdavi@sfu.ca}

\author{Daniel Cohen-Or}
\affiliation{
  \institution{Tel Aviv University}
  \country{Israel}
}
\email{cohenor@gmail.com}

\author{Yiorgos Chrysanthou}
\affiliation{
  \institution{University of Cyprus,}
  \institution{CYENS Center Of Excellence}
  \country{Cyprus}
}
\email{chrysanthou.yiorgos@ucy.ac.cy}

\newif\ifshowsuppfigs
\showsuppfigstrue 

\begin{document}

\begin{abstract}
  Motion-centric video editing remains difficult for large generative video models, which often respond well to appearance changes but struggle to produce specific, localized actions or state transitions in an existing clip. We introduce Sound Sparks Motion, a training-free framework that enables motion editing in an audio-visual video generation model by tuning its internal multimodal conditioning signals at test time. Rather than modifying model weights, our method tunes only two lightweight variables: an audio latent derived from the source video and a residual perturbation in the text-conditioning. We find that this combination can encourage motion edits that the underlying model often struggles to realize under prompt-only control.
Since there is no direct way to evaluate temporal alignment between text and motion, we guide the tuning process using a vision–language model that provides feedback indicating whether the intended motion appears in the generated video. This simple supervision yields an effective semantic objective for motion editing, while regularization and perceptual-temporal constraints help preserve content and visual quality. Beyond per-video tuning, we show that the learned latent controls are transferable across videos, suggesting that they capture reusable motion-edit directions rather than overfitting to a single example.
Our results highlight multimodal conditioning tuning, particularly through the audio pathway, as a promising direction for motion-aware video editing, and suggest that test-time tuning can serve as a lightweight probing mechanism that helps reveal latent motion controls embedded in the model’s multimodal conditioning. Code and data are available via our project page: \url{https://amirhossein-razlighi.github.io/Sound_Sparks_Motion/}.

\end{abstract}

\begin{teaserfigure}
\centering
  \includegraphics[width=\textwidth]{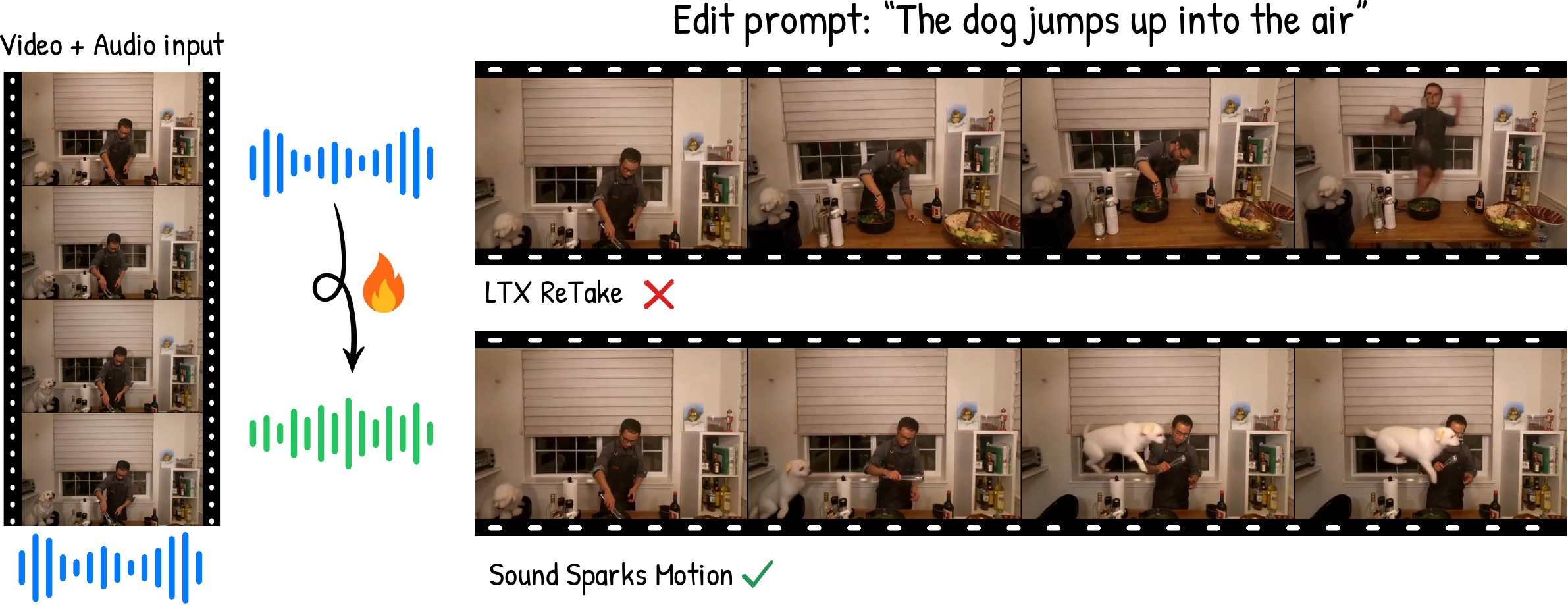}
  \caption{Given a source video and an edit prompt, our method tunes the audio latent as a conditioning parameter to the LTX video model, such that the desired motion edit is realized in the edited video.}
  \Description{}
  \label{fig:teaser}
\end{teaserfigure}

\maketitle

\section{Introduction}
\label{sec:intro}

Video generation models usually do not deliver fully satisfactory results in a single pass. In practice, creating the desired outputs often involves an iterative process in which both visual appearance and motion dynamics are gradually refined. Users frequently need to adjust temporal consistency, object movement, and scene composition across multiple iterations to achieve coherent and realistic results. Most current video generation systems are primarily controlled through text-based prompts. While text provides a convenient and flexible interface, it is often insufficient for precisely specifying motion, especially in audio-visual generative models. In such models, motion is tightly entangled with both visual content and audio cues, making it difficult to isolate and control through language alone. As a result, relying solely on text limits the ability to guide motion effectively.
\begin{figure}[t]
    \centering
    \includegraphics[width=\linewidth]{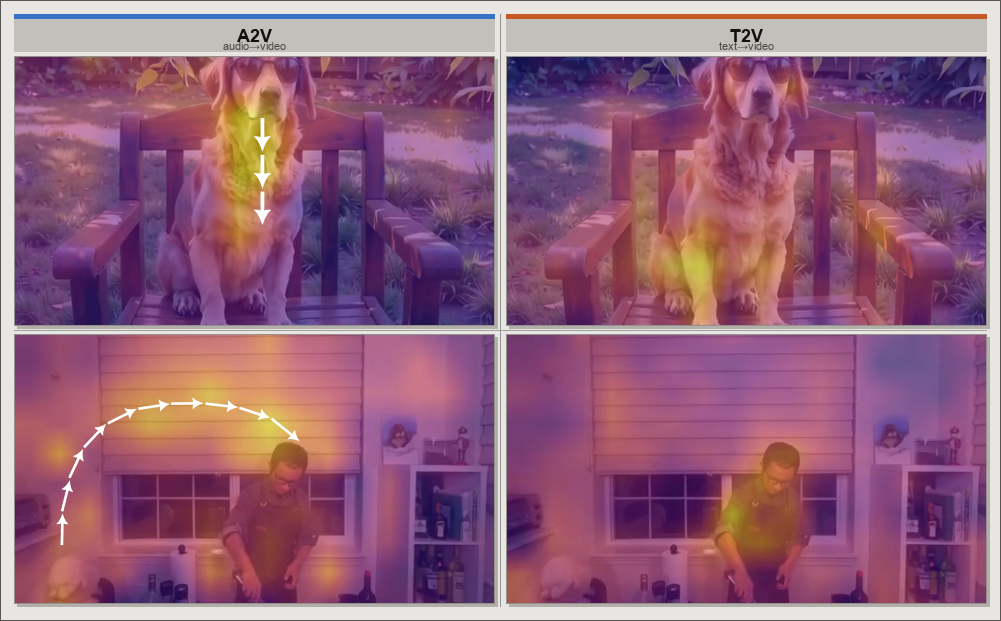}
    \vspace{-0.8em}
    \caption{
    A visualization of average attention heat maps for tuned audio and text components, plotted on top of the first frame of the video. \textbf{"A2V"} shows "audio-to-video" cross attention and \textbf{"T2V"} shows "text-to-video" cross attention. Notice how text focuses more on the semantics and objects, while audio is focusing more on motion trajectory (shown with white arrows). Here the motion is yawning (top) and jumping (bottom) for the dogs.
    }
    \label{fig:attention}
    \vspace{-1.0em}
\end{figure}

In this setting, once a video is generated, performing targeted edits—such as modifying existing motions or introducing new actions—remains challenging. Many recent approaches follow a paradigm in which selected segments of the video are re-generated conditioned on the input video and a text prompt. This form of conditional re-synthesis, sometimes referred to as \emph{retake-style} editing, allows the model to preserve parts of the original content while re-generating others. However, when attempting to induce specific motion changes through prompt adjustments alone, the model often struggles to realize the intended behavior and instead reproduces the original motion (see Fig.~\ref{fig:teaser}).

In this work, we revisit the role of multimodal conditioning in motion generation and show that audio plays a key role in enabling motion. We observe that motion is strongly coupled with audio signals in the model’s internal representation, even in cases where no natural sound is explicitly associated with the action (see Fig.~\ref{fig:attention}). Building on this, we propose to tune the audio conditioning of a video generation model at test time in order to encourage the desired motion. In addition, we introduce a lightweight residual adjustment in the text-conditioning space, allowing the model to better align with the intended edit without modifying the prompt itself (Fig~\ref{fig:text_vs_both}).

To guide this process, we employ a vision-language model that provides binary feedback indicating whether the intended motion or state change appears in the generated video. This feedback defines a simple semantic objective for test-time tuning, while regularization and perceptual-temporal constraints help preserve content and visual quality. Importantly, the tuning operates over a small set of conditioning variables, leaving all model parameters unchanged.

We instantiate our approach using LTX, an open-source audio-visual video generation model that exposes its internal multimodal conditioning signals. This accessibility enables us to probe the role of audio and text conditioning in motion generation. While our implementation utilizes LTX,  audio and motion are naturally entangled, and this coupling likely arises in other joint audio-visual architectures with a similar conditioning structure. Our framework should therefore be applicable to such models.


Beyond per-video tuning, we observe that the learned conditioning adjustments can transfer across videos, suggesting that they capture reusable motion-edit directions rather than overfitting to a single instance. Taken together, these findings suggest that test-time tuning can serve as a lightweight probing mechanism that helps reveal latent motion controls embedded in the model’s multimodal conditioning.

\section{Related Work}
\label{sec:related}

\noindent\textbf{Text-guided video editing.}
Recent advances in generative visual content creation have unlocked powerful video editing capabilities.
Early works approached video editing by extending text-to-image diffusion models to the video domain, adapting image-editing techniques such as inversion, attention control, and feature propagation to achieve temporal consistency and coherent editing across frames \cite{tokenflow2023, cong2023flatten, qi2023fatezero, liu2023videop2p, wang2025videodirector}.

With the emergence of specialized video generative models~\cite{opensora, wan2025, kong2024hunyuanvideo, ltx2}, subsequent editing methods increasingly built on these native backbones. Some of these approaches operate purely at inference time, leveraging pretrained video generators for tuning-free editing without additional model training~\cite{bai2024uniedit, ku2024anyv2v, xu2025freevis}.
In parallel, another line of work improves editability and controllable generation through model-side adaptation, including finetuning~\cite{molad2023dreamix}, richer conditioning architectures and adapter modules~\cite{vace, Kiwi-Edit2026, ye2025stylemaster}, lightweight LoRA~\cite{hu2021lora} for extra conditioning~\cite{benyosef2026avcontrol}, first-frame edit propagation~\cite{polaczek2024synclora, gao2026controllable}, and fully trained instruction-guided editors~\cite{qin2023instructvid2vid, lucyedit2025}.

\begin{figure*}[t]
    \centering
    \includegraphics[width=\textwidth]{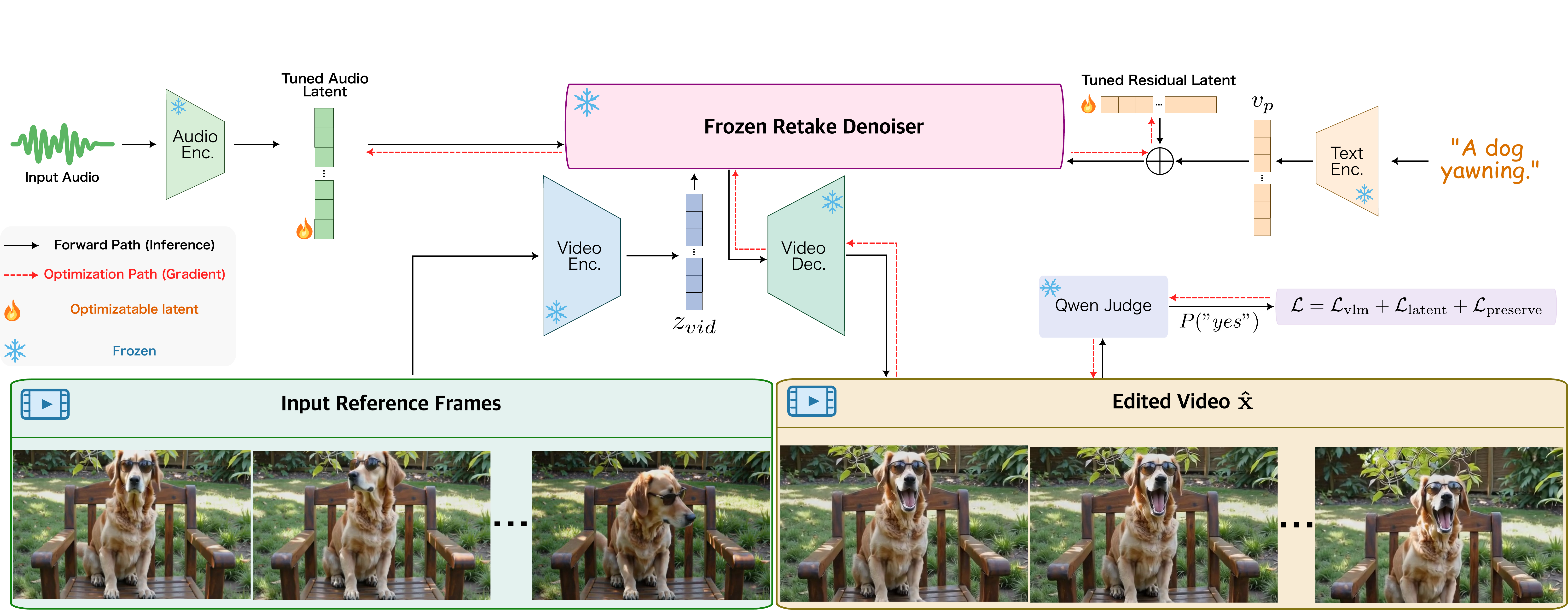}
    \vspace{-0.8em}
    \caption{
    \textbf{Overview of \paper{}.}
    Given source audio, reference video frames, and an edit prompt, our method keeps LTX Retake, Text Encoder, Qwen, and the VAE modules frozen, and optimizes only the audio latent and a residual text-conditioning vector. The edited video is temporally sampled and scored by Qwen2.5-VL through a binary motion question, producing a test-time loss that updates only the learnable latents while preserving the frozen generator, making it capable of complex motion edits.
    }
    \label{fig:arch}
    \vspace{-1.0em}
\end{figure*}

Despite this progress, current text-guided video editing methods are more effective at changes in appearance, style, and semantic attributes than at motion-centric edits. In contrast to appearance edits, motion edits are difficult both to control at inference time and to supervise during training, as they require temporally grounded instructions and high-quality source-target video pairs that modify motion while preserving content. Our work addresses this gap by showing that, in joint audio-video models~\cite{ltx2}, the audio pathway can serve as an internal source of temporal control for motion-centric editing.

\noindent\textbf{Motion control and editing.}
Motion in videos arises from both camera and object movement. 
For camera motion, several works condition video generation on explicit camera poses or trajectories~\cite{bahmani2025ac3d, he2024cameractrl, he2025cameractrlii, bahmani2025vd3d, ren2025gen3c}. Related methods instead re-render an existing video under a new camera trajectory~\cite{li2026rerope, bai2025recammaster}.

For object motion, some methods transfer motion from a reference video by manipulating attention~\cite{pondaven2025ditflow}, and positional encoding~\cite{gokmen2025ropecrafttrainingfreemotiontransfer}, while others use explicit controls such as trajectories~\cite{geng2024motionprompting, namekata2024sgi2v, ma2023trailblazer}, flow fields~\cite{burgert2025gowiththeflow}, and physical force~\cite{gillman2025forcepromptingvideogeneration, gillman2026goalforceteachingvideo} for motion-controllable video generation. Related work also studies trajectory-guided inbetweening between endpoint frames~\cite{framer, multicoin}. 
Closer to our setting, motion editing methods modify the dynamics of an existing video using trajectories~\cite{mou2024revideo, burgert2025motionv2veditingmotionvideo}, or anchor-frame paths~\cite{phung2026trace, lee2025editbytrack} while preserving scene content.

In contrast, our method does not require external motion annotations or user-specified controls. While recent audio-conditioned editing approaches such as Just-Dub-It~\cite{chen2026justdubit} use target speech to drive synchronized facial motion for dubbing, we instead study the interaction between audio and video conditioning in a joint audio-video model, showing that the internal audio pathway can serve as a latent source of temporal control for general text-guided motion editing.

\enlargethispage{1\baselineskip}
\noindent\textbf{Reward/VLM-guided prompt alignment.}
Recent work has aligned diffusion models with human preferences and high-level semantic objectives by using learned reward models and VLM feedback, rather than relying solely on paired supervision. On the image side, prior works have used reward models and preference optimization for text-to-image generation~\cite{xu2023imagereward, prabhudesai2023aligning, wallace2023diffusion}.  EditReward~\cite{wu2025editreward} uses reward optimization for instruction-guided image editing. FireEdit uses region-aware VLM feedback for fine-grained image editing~\cite{gan2025fireedit}, and ParetoSlider~\cite{golan2025paretoSlider} studied post-training for continuous appearance control over competing reward objectives.

On the video side, reward-based alignment has been used to improve prompt faithfulness and generation quality through pretrained reward models, learned critics from human feedback, and preference-based post-training~\cite{prabhudesai2024videodiffusionalignmentreward, wang2024lift, liu2025improving}. More closely related to our setting, VIVA~\cite{cong2025viva} uses VLM-guided reward optimization for instruction-based video editing, and Diffusion-DRF~\cite{wang2026diffusiondrfdifferentiablerewardflow} shows that frozen off-the-shelf VLMs can act as critics for video diffusion optimization. In contrast, our method does not train a separate reward model or post-train the video generator; instead, we use a frozen VLM only at test time to guide per-video optimization of audio latents for motion-centric editing.

\section{Method}\label{sec:method}

Given an input video $\mathbf{x} = \{x_t\}_{t=1}^{T}$, where $x_t$ denotes the RGB frame at time $t$, and a natural-language edit instruction $p$, our goal is to generate an edited video $\hat{\mathbf{x}} = \{\hat{x}_t\}_{t=1}^{T}$ that satisfies the instruction while preserving the identity, scene layout, and visual content of the source. We focus on \emph{motion-centric} edits, where the desired change is primarily temporal rather than appearance-based, e.g., making a subject jump, yawn, bloom, or perform a specified action.

We assume access to a pretrained video-audio generation model $\mathcal{G}_{\theta}$ with frozen parameters $\theta$; in our implementation, we use LTX2~\cite{ltx2}. In the standard setting, the model produces an edited video from a source video segment and an edit prompt,
\begin{equation}
    \hat{\mathbf{x}} = \mathcal{G}_{\theta}(\mathbf{x}, p).
\end{equation}
However, editing via prompt-only conditioning often fails to realize the requested motion or state change. Therefore, we intend to design a training-free procedure that improves motion controllability without updating $\theta$.

At a high level, the desired edit should satisfy the prompt ($\mathcal{L}_{\mathrm{edit}}$) while preserving the source ($ \mathcal{L}_{\mathrm{preserve}}$):
\begin{equation}
    \hat{\mathbf{x}}^{*}
    =
    \arg\min_{\hat{\mathbf{x}}}
    \;
    \mathcal{L}_{\mathrm{edit}}(\hat{\mathbf{x}}, p)
    +
    \lambda
    \mathcal{L}_{\mathrm{preserve}}(\hat{\mathbf{x}}, \mathbf{x}).
\end{equation}
The next sections describe how we realize this objective by tuning internal conditioning latents of the frozen editor at test time.

\subsection{Method Overview}
\label{sec:method_overview}

Our method turns a frozen retake-style video editor into a motion-aware editor through test-time latent tuning, as illustrated in Fig.~\ref{fig:arch}. Given a source segment $\mathbf{x}_{i:j}$ and an edit prompt $p$, we search the editor's internal multimodal conditioning space for controls that elicit the desired action, while keeping all model weights fixed.

We tune two lightweight variables: an audio latent $\mathbf{\alpha}$ initialized from the source audio $\mathbf{\alpha}_0$, and a residual text-conditioning variable $\Delta\mathbf{v}$ added to the video-text representation of the edit prompt. Together, they define the modified conditioning state
\begin{equation}
    \mathbf{c}
    =
    \left(
        \mathbf{\alpha},
        \mathbf{v}_{p} + \Delta\mathbf{v}
    \right),
\end{equation}
where $\mathbf{v}_{p}$ is the original prompt-conditioned video representation. Only $\mathbf{\alpha}$ and $\Delta\mathbf{v}$ are optimized; the video editor, text encoder, audio encoder, and all auxiliary models remain frozen.

We use different parameterizations for the two conditioning branches. For text, we optimize a residual because the prompt representation depends on the edit instruction and may vary with prompt content and length; adding $\Delta\mathbf{v}$ keeps the optimized representation explicitly anchored to the original prompt semantics. In contrast, the audio-conditioning latent is tied to the fixed source clip and duration, and is already a compact continuous conditioning variable. We therefore optimize $\mathbf{\alpha}$ directly, while regularizing it toward $\mathbf{\alpha}_0$, which provides the same anchoring effect while allowing more direct control over the temporally structured audio pathway.

\begin{table}[t]
    \centering
    \setlength{\tabcolsep}{5pt}
    \renewcommand{\arraystretch}{1.10}
    \caption{%
        \textbf{Ablation of guidance modality (VLM evaluation, 10 scenarios).}
        Scores on a 1--10 scale.
        \textit{Alignment}\ reflects semantic fidelity to the text prompt (primary role of \textbf{text} guidance).
        \textit{Naturalness}\ reflects motion smoothness and temporal coherence (primary role of \textbf{audio} guidance).
        \textit{Preservation}\ measures source-identity retention.
    }
    \label{tab:ablation_modality}
    \begin{tabularx}{\columnwidth}{@{}X ccc@{}}
        \toprule
        \textbf{Metric}
          & \textbf{Text}
          & \textbf{Audio}
          & \textbf{Both} \\
        \midrule
        Alignment$\uparrow$~{\scriptsize\textit{(text role)}}
          & 7.3 & 7.1 & \textbf{7.5} \\
        Naturalness$\uparrow$~{\scriptsize\textit{(audio role)}}
          & 7.0 & 7.3 & \textbf{7.9} \\
        \midrule
        Preservation$\uparrow$
          & 7.6 & 7.8 & \textbf{8.4} \\
        Overall$\uparrow$
          & 7.25 & 7.24 & \textbf{7.72} \\
        \bottomrule
    \end{tabularx}
\end{table} 

\subsection{Latent Motion Control in LTX Retake}
\label{sec:method_latent_control}

We instantiate $\mathcal{G}_{\theta}$ with LTX2 Retake, a pretrained video-audio editing pipeline that regenerates a selected temporal region of a source video conditioned on text. We use it as a frozen backbone and tune only selected conditioning variables for each edit. Since LTX is currently the only publicly available audio-visual generation model with editing capabilities, our implementation is built for it.

Given an editable segment $\mathbf{x}_{i:j}$ and prompt $p$, LTX Retake constructs a multimodal conditioning state from source video, source audio, and text. We keep the source video conditioning fixed and tune two components: the audio-conditioning latent and the text-video conditioning representation. Let $\mathbf{\alpha}_0$ be the source audio latent and $\mathbf{v}_p$ the video-conditioning component of the text representation. The prompt-only baseline is
\begin{equation}
    \hat{\mathbf{x}}_{\mathrm{base}}
    =
    \mathcal{G}_{\theta}
    \left(
        \mathbf{x}_{i:j};
        \mathbf{\alpha}_0,
        \mathbf{v}_p
    \right).
\end{equation}
Our method instead generates
\begin{equation}
    \hat{\mathbf{x}}
    =
    \mathcal{G}_{\theta}
    \left(
        \mathbf{x}_{i:j};
        \mathbf{\alpha},
        \mathbf{v}_p + \Delta \mathbf{v}
    \right),
    \qquad
    \theta \ \text{fixed},
\end{equation}
with $\mathbf{\alpha}$ initialized as $\mathbf{\alpha}_0$ and $\Delta\mathbf{v}$ initialized as zero.

\noindent\textbf{Audio-Conditioning Latent Tuning.}
The first control variable is the audio-conditioning latent $\mathbf{\alpha}$. Although our target edits are evaluated visually, the audio pathway of a joint video-audio generator provides a temporally structured conditioning signal aligned with the generated video. Attention visualizations in Fig.~\ref{fig:attention} suggest that the audio pathway is spatially and temporally correlated with localized motion regions. We exploit this pathway as a latent handle for motion control.

We initialize $\mathbf{\alpha}$ from the source video's encoded audio latent $\mathbf{\alpha}_0$. This choice keeps the optimization close to the original source-conditioned generation and avoids searching from an arbitrary latent code. During test-time editing, gradients update $\mathbf{\alpha}$ so that the generated video better expresses the motion or state change requested by the prompt. The audio latent is therefore used as a controllable temporal conditioning variable, not as a mechanism for audio editing itself.

\noindent\textbf{Residual Text-Conditioning Tuning.}
The second control variable is a residual perturbation $\Delta \mathbf{v}$ in the text encoder's video-conditioning space. We keep the edit prompt tokens fixed and do not perform prompt rewriting. Given the edit prompt $p$, the frozen text encoder produces a video-conditioning representation $\mathbf{v}_p$. We add a learnable residual to this representation,
\begin{equation}
    \tilde{\mathbf{v}}_p
    =
    \mathbf{v}_p + \Delta \mathbf{v},
\end{equation}
and use $\tilde{\mathbf{v}}_p$ as the text-conditioning input to the frozen video editor.

\begin{figure}[t]
    \centering
    \includegraphics[width=\linewidth]{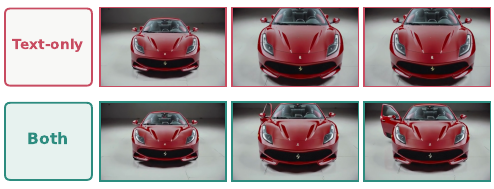}
    \vspace{-0.8em}
    \caption{
    A comparison between text-only residual tuning and tuning text and audio at the same time on \emph{"a red car door opens."} shows that in some cases, the model fails to capture all the required information to do hard motions while keeping the main subjects in the scene unchanged in text-only mode.}
    \label{fig:text_vs_both}
    \vspace{-1.0em}
\end{figure}

The audio latent provides a temporally aligned handle on generation, while the text residual adjusts semantic conditioning without modifying the prompt; since $\Delta \mathbf{v}$ is initialized to zero, it only deviates from the original prompt when doing so improves the motion edit objective. As shown in Fig.~\ref{fig:text_vs_both} and Tab.~\ref{tab:ablation_modality}, the two controls are complementary---text-residual tuning improves semantic alignment while audio-latent tuning improves motion realization and quality---and joint tuning outperforms either alone across all metrics.

\subsection{VLM-Guided Motion Supervision}
\label{sec:method_vlm_supervision}

\begin{figure}[t]
    \centering
    \includegraphics[width=\linewidth]{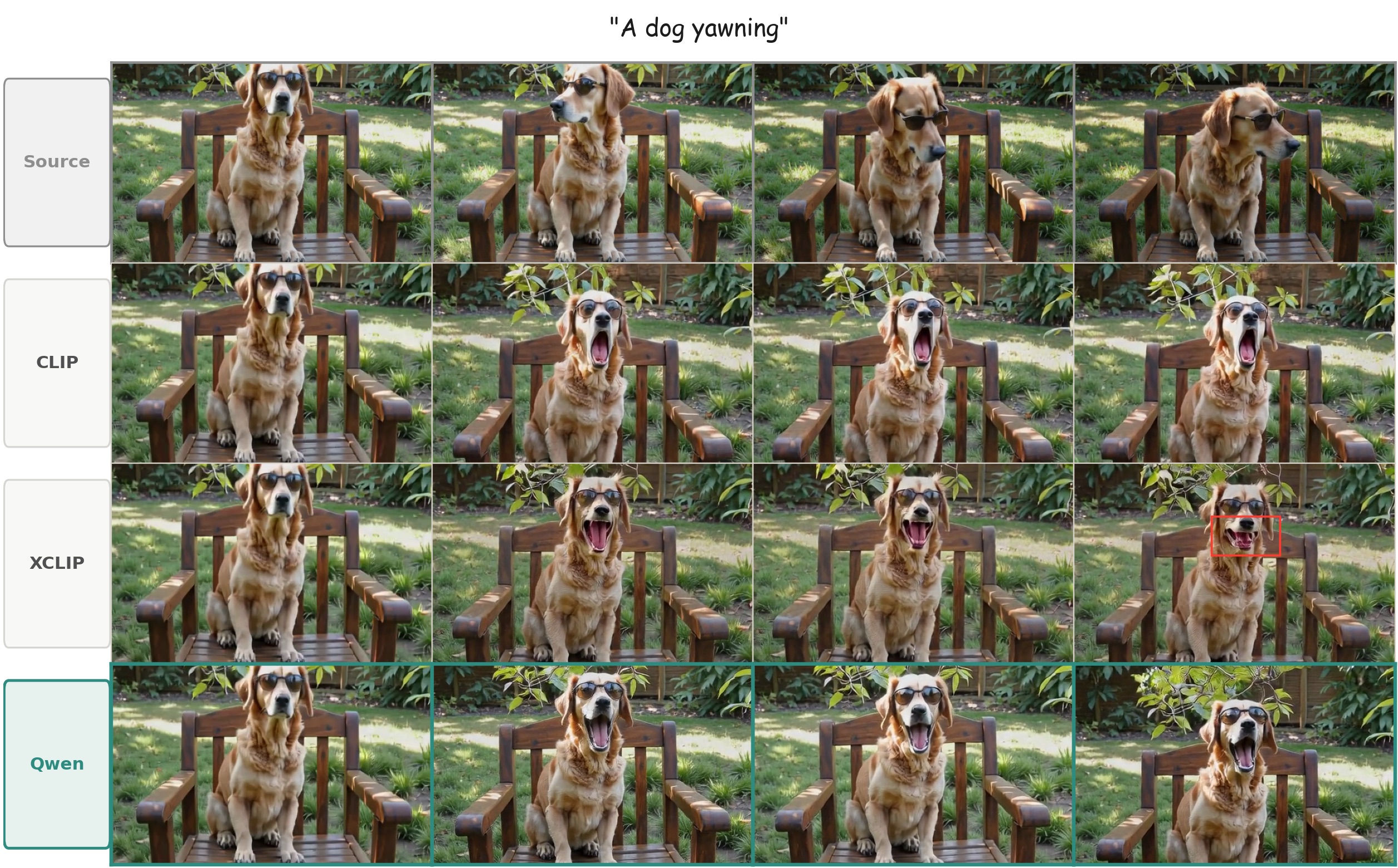}
    \vspace{-1.5em}
    \caption{
    \textbf{Effect of supervision method.}
        CLIP~\cite{clip} guidance primarily optimizes frame-level semantic appearance rather than temporal action, leading to a mostly static yawning state. XCLIP~\cite{xclip} provides limited temporal awareness, but still introduces global appearance drift and localized artifacts, highlighted by the red box. Qwen-based supervision better captures whether the requested motion unfolds over time, producing a more coherent yawning edit.
    }
    \label{fig:ablation_scorer}
    \vspace{-1.0em}
\end{figure}

Motion edits are difficult to supervise with pixel-level, perceptual, or temporal smoothness losses alone, since the desired change is specified semantically by the edit prompt. Frame-level image-text models such as CLIP~\cite{clip} are also insufficient, as they score sampled frames independently and often miss temporal actions or state transitions (see Fig.~\ref{fig:ablation_scorer}). We therefore use a frozen vision-language model as an edit critic. As shown in Fig.~\ref{fig:arch}, this supervision forms a closed test-time optimization loop: generated frames are scored by Qwen, and gradients update only the learnable conditioning variables.

Given the edit prompt $p$, we form a binary question
\begin{equation}
\begin{aligned}
    q(p) =\;&
    \text{``Does this video clearly show the action or state change} \\
    &\text{described by the edit prompt: } p \text{?''}
\end{aligned}
\end{equation}
and evaluate the generated video $\hat{\mathbf{x}}$ with Qwen2.5-VL~\cite{qwen}. The model is kept frozen and is used only to provide a differentiable supervision signal.

Let $\ell_{\mathrm{yes}}$ and $\ell_{\mathrm{no}}$ be the logits assigned by Qwen to the answer tokens \texttt{yes} and \texttt{no} at the final response position. We convert them into a binary probability distribution,
\begin{equation}
    P_{\mathrm{yes}}
    =
    \frac{
        \exp(\ell_{\mathrm{yes}})
    }{
        \exp(\ell_{\mathrm{yes}})
        +
        \exp(\ell_{\mathrm{no}})
    },
\end{equation}
and define the VLM edit loss as
\begin{equation}
    \mathcal{L}_{\mathrm{vlm}}
    =
    -\log
    \left(
        P_{\mathrm{yes}} + \epsilon
    \right),
\end{equation}
where $\epsilon$ is a small constant for numerical stability. Minimizing this loss increases the likelihood that the generated video is judged to contain the requested motion or state change.

For efficiency, Qwen is applied to a temporally sampled subset of generated frames rather than the full video. We use either \textbf{uniform sampling} across the clip or a \textbf{midpoint-biased normal sampling} schedule, which provides denser supervision around the central editable region while still covering the beginning and end of the video. In supplementary material Fig.S3, we show that the sampling strategy can also act as an optional lightweight control over the temporal emphasis and apparent speed of the optimized motion.

\subsection{Regularized Test-Time Tuning}
\label{sec:method_objective}

\begin{figure}[t]
    \centering
    \includegraphics[width=0.95\linewidth]{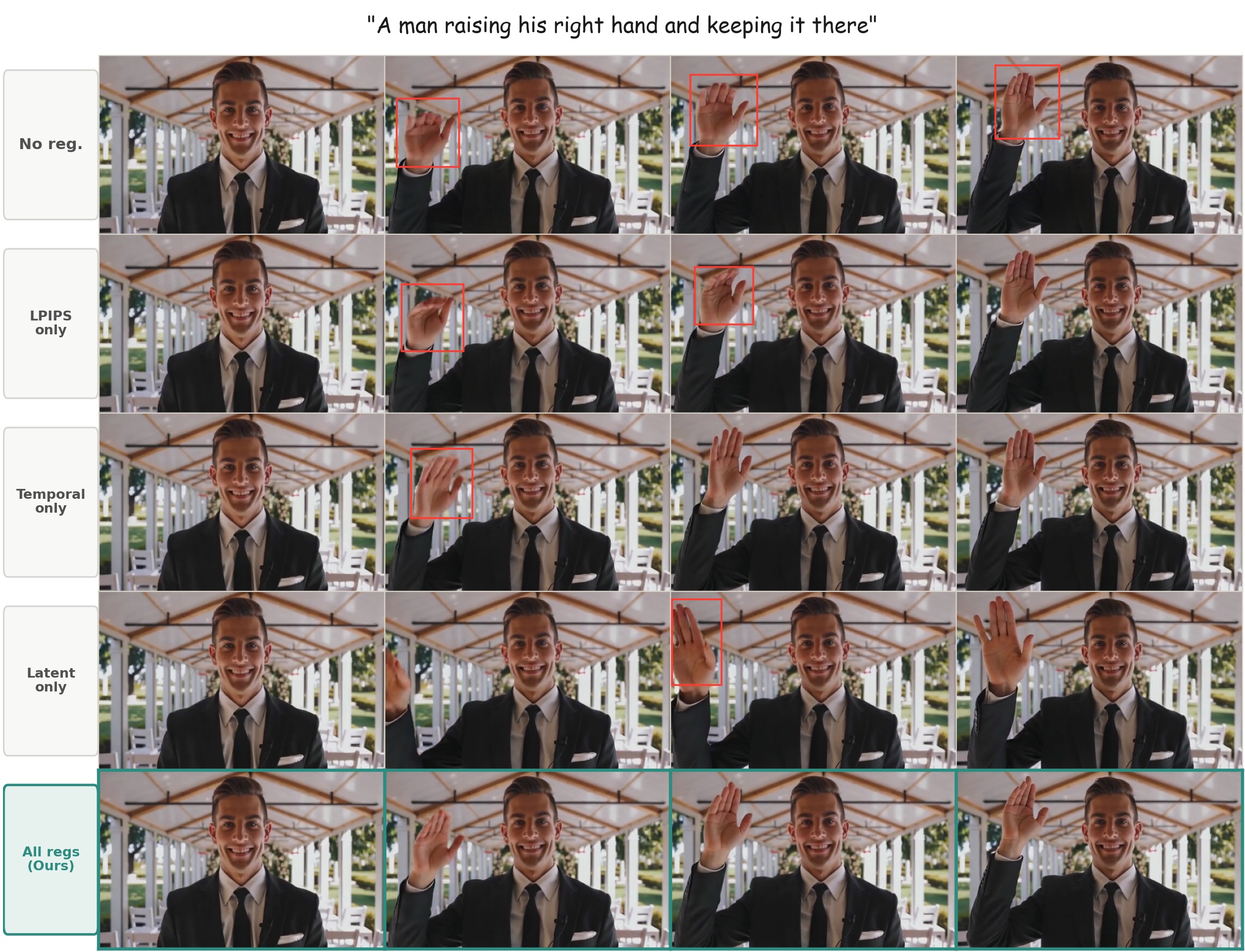}
    \vspace{-0.8em}
    \caption{
    \textbf{Effect of regularization.}
    Without regularization, VLM-guided tuning produces artifacts and abrupt motion, with representative failures highlighted in red. Latent regularization reduces drift, LPIPS improves frame-level fidelity, and temporal regularization improves motion smoothness. Combining all terms yields the most coherent and stable result.
}
    \label{fig:ablation_reg}
    \vspace{-1.0em}
\end{figure}

The VLM loss encourages semantic edit, but alone may cause latent drift, visual artifacts, or adversarial solutions that exploit the critic (see Fig.~\ref{fig:ablation_reg}). We therefore regularize the tuned variables and add optional perceptual-temporal preservation terms.
We keep the optimized audio latent close to the source audio latent $\mathbf{\alpha}_0$ and penalize large residual text-conditioning updates, i.e., large values of $\Delta\mathbf{v}$:

\begin{equation}
    \mathcal{L}_{\mathrm{latent}}
    =
    \lambda_\alpha
    \left\|
        \mathbf{\alpha} - \mathbf{\alpha}_0
    \right\|_2^2
    +
    \lambda_v
    \left\|
        \Delta \mathbf{v}
    \right\|_2^2 .
\end{equation}
This anchors the search near the original source-conditioned retake behavior while still allowing the latents to move toward the requested edit. 

For visual preservation, we compare the optimized video to the prompt-only retake baseline rather than to the raw input video. Let $\hat{\mathbf{x}}_{\mathrm{base}}$ denote this baseline and $\hat{\mathbf{x}}$ the current optimized output:
\begin{equation}
    \mathcal{L}_{\mathrm{preserve}}
    =
    \lambda_{\mathrm{lpips}}
    \mathcal{L}_{\mathrm{LPIPS}}
    (
        \hat{\mathbf{x}},
        \hat{\mathbf{x}}_{\mathrm{base}}
    )
    +
    \lambda_{\mathrm{temp}}
    \mathcal{L}_{\mathrm{temp}}
    (
        \hat{\mathbf{x}},
        \hat{\mathbf{x}}_{\mathrm{base}}
    ).
\end{equation}
The LPIPS~\cite{lpips} term discourages unnecessary perceptual deviation from the prompt-only retake result. The temporal term measures LPIPS between consecutive frames and penalizes excessive temporal variations beyond that of the baseline, reducing flicker without suppressing natural motion.

For temporal preservation, we define
\begin{equation}
    \mathcal{L}_{\mathrm{temp}}
    =
    \max
    \left(
        0,
        D_{\mathrm{temp}}(\hat{\mathbf{x}})
        -
        D_{\mathrm{temp}}(\hat{\mathbf{x}}_{\mathrm{base}})
    \right),
\end{equation}
where $D_{\mathrm{temp}}$ is the mean LPIPS distance between sampled consecutive frame pairs. This penalizes flicker and abrupt changes while allowing motion already present in the baseline.

The final test-time objective is
\begin{equation}
    \mathcal{L}
    =
    \mathcal{L}_{\mathrm{vlm}}
    +
    \mathcal{L}_{\mathrm{latent}}
    +
    \mathcal{L}_{\mathrm{preserve}} .
\end{equation}
We optimize this objective only with respect to $\mathbf{\alpha}$ and $\Delta\mathbf{v}$; all pretrained model parameters remain fixed.

\section{Experiments}
\label{sec:experiments}

\subsection{Experimental Setup}
\label{sec:exp_setup}

We evaluate \paper{} on 25 short motion-centric editing tasks, each consisting of a source video and an edit prompt describing a temporal change, e.g., jumping, yawning, wing opening, blooming, door opening, etc. Since existing video-editing benchmarks do not specifically target prompt-driven actions and state changes, we curate a controlled benchmark with 20 AI-generated and 5 real videos. The generated clips provide clean, unambiguous edit settings, while the real clips test generalization beyond generated inputs. We run our method on all 25 tasks and compare against all baselines on a representative subset of 12 tasks, chosen for diversity and due to baseline running costs.

Unless otherwise stated, videos are edited at 512$\times$320 resolution, 25 FPS, and 89 frames. The number of preserved conditioning frames varies from 3 to 25 and is reported per experiment, depending on the edit complexity and the desired strength of source preservation. For baselines, we use the same source videos, prompts, resolution, and conditioning prefix whenever supported. If a baseline method requires extra inputs, such as masks or edited reference frames, we follow its official setup and provide the required inputs.

\begin{table}[t]
    \centering
    \setlength{\tabcolsep}{3.0pt}
    \renewcommand{\arraystretch}{1.08}
    \caption{
    \textbf{Human evaluation across 12 scenarios.}
    Win Rate is the fraction of first-place selections; Top-3 Rate is the fraction of trials where the method appears in the top three; AVG Score assigns 3/2/1 points to first/second/third place; Task Achieved is the fraction of user achievement checks.
    }
    \label{tab:quantitative_human_survey}
    \begin{tabularx}{\columnwidth}{@{}Xcccc@{}}
        \toprule
        \textbf{Method} 
        & \textbf{Win}$\uparrow$ 
        & \textbf{Top-3}$\uparrow$ 
        & \textbf{Score}$\uparrow$ 
        & \textbf{Ach.}$\uparrow$ \\
        \midrule
        VACE-Wan           & 0.0\%  & 8.3\% & 0.10 & 10.8\% \\
        VACE-LTX           & 0.0\%  & 1.2\%  & 0.02 & 2.5\% \\
        Kiwi-Edit          & 0.0\%  & 10.7\% & 0.12 & 10.8\% \\
        UniEdit            & 0.8\%  & 4.0\%  & 0.06 & 5.7\% \\
        LTX2.3 Retake      & 10.3\%  & 63.1\% & 1.08 & 39.2\% \\
        Runway-Aleph       & 11.1\% & 58.3\% & 0.98 & 41.8\% \\
        LoRA-Edit          & 21.4\% & 59.1\% & 1.28 & 48.7\% \\
        \textbf{\paper{}}  & \textbf{56.3\%} & \textbf{95.2\%} & \textbf{2.37} & \textbf{79.1\%} \\

        \bottomrule
    \end{tabularx}
\end{table}

\noindent\textbf{Baselines.}
We compare with text-guided and general-purpose video editing systems: \textbf{LTX2.3 Retake}, our prompt-only backbone baseline without latent tuning; \textbf{UniEdit}, a tuning-free inversion-then-generation video editing framework; \textbf{Kiwi-Edit}, an instruction- and reference-guided video editor used in its instruction-guided mode; \textbf{VACE}, evaluated in video-to-video mode with both Wan2.1-1.3B and LTX-Video-2B~\cite{ltx_video} backbones; \textbf{LoRA-Edit}~\cite{lora_edit}, a first-frame-guided editing method for which we provide the required edited first-frame and an additional edited final-frame for better results; and \textbf{Runway-Aleph}~\cite{runway2025aleph}, a commercial general-purpose video editing model.

\noindent\textbf{Evaluation Protocol.}
We use both VLM-based evaluation and a human user study. For automatic evaluation, GPT-5.5-Thinking receives the source video, edit prompt, and all candidate edits, and scores each result for task success, edit alignment (EA), motion quality (MQ), source preservation (SP), and visual quality (VQ). Since our benchmark targets motion editing, we assign slightly larger weights to edit alignment and motion quality, while still giving substantial weight to preservation and visual fidelity:

\begin{equation}
    S_{\mathrm{vlm}}
    =
    0.35 S_{\mathrm{EA}}
    +
    0.25 S_{\mathrm{MQ}}
    +
    0.20 S_{\mathrm{SP}}
    +
    0.20 S_{\mathrm{VQ}} .
\end{equation}
Task success (TS) is the percentage of scenarios judged to clearly achieve the requested motion or state change.

For human evaluation, 21 participants evaluate the same 12 scenarios in an anonymous randomized survey. Within each scenario, participants see the source video, edit prompt, and eight anonymized results in randomized order, corresponding to our method and baselines. Participants select their top three results and mark videos where the edit is clearly achieved. We report \textbf{Win Rate}, the fraction of first-place selections; \textbf{Top-3 Rate}, the fraction of trials in which a method appears in a participant's top three; \textbf{AVG Score}, computed with 3, 2, and 1 points for first, second, and third place; and \textbf{Task Achieved}, the fraction of achieved check boxes selected by users.

\begin{figure*}[t]
    \centering
    \includegraphics[width=0.97\textwidth]{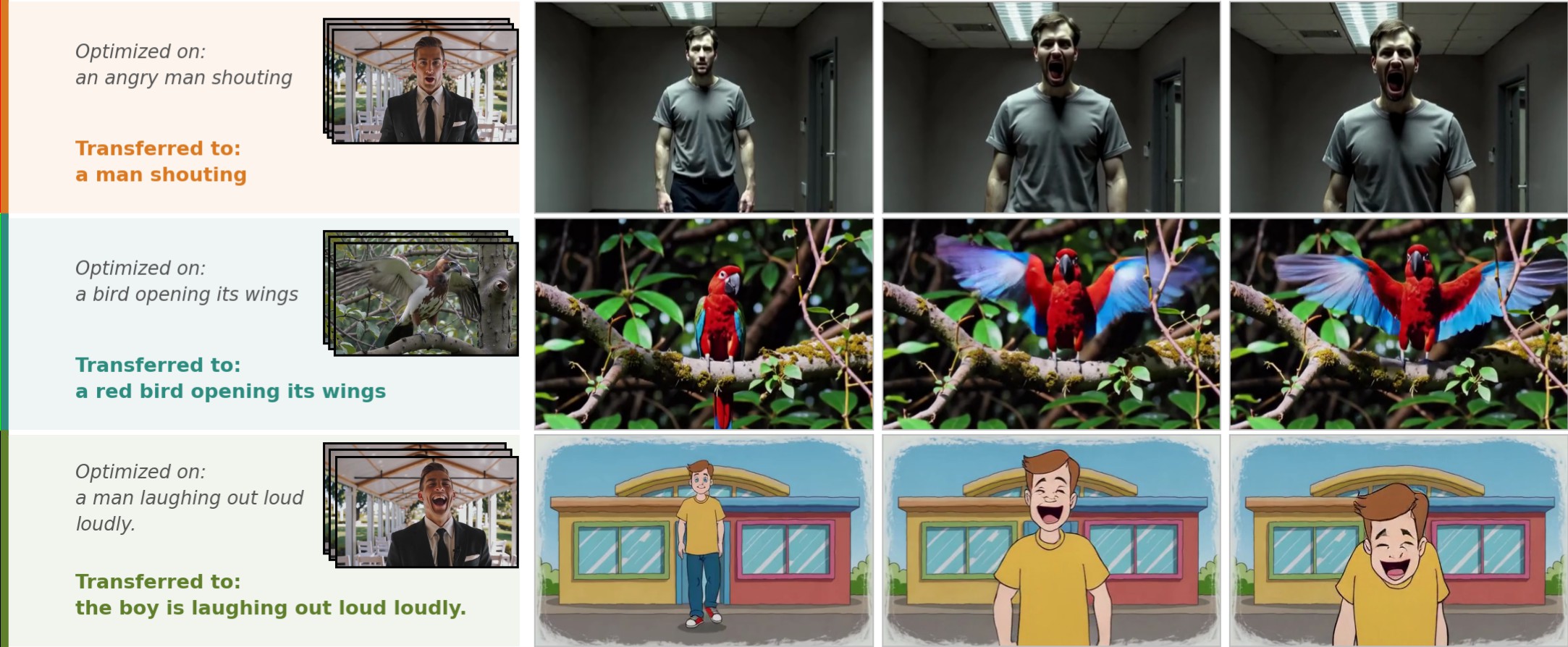}
    \vspace{-1.0em}
    \caption{
    \textbf{Transferability of optimized latents.}
    Learned audio and residual text latents are reused on related target prompts without additional tuning, transferring motions such as yawning, shouting, and wing opening across different subjects and scenes. Note that in the last row although the latents are initially learned for a realistic video, they can be transferred to a cartoonish video, which shows the generality of the learned latents.
    }
    \label{fig:transferability_exp}
    \vspace{-1.0em}
\end{figure*}

\begin{table}[t]
    \centering
    \setlength{\tabcolsep}{3.0pt}
    \renewcommand{\arraystretch}{1.08}
    \caption{
    \textbf{GPT-5.5-Thinking evaluation across 12 scenarios.}
    TS is task success rate; VLM Score is the weighted 0--10 score; EA and MQ denote edit alignment and motion quality; VLM Rank is the average comparative rank, lower is better.
    }
    \label{tab:quantitative_vlm}
    \begin{tabularx}{\columnwidth}{@{}Xccccc@{}}
        \toprule
        \textbf{Method} 
        & \textbf{TS}$\uparrow$ 
        & \textbf{VLM}$\uparrow$ 
        & \textbf{EA}$\uparrow$ 
        & \textbf{MQ}$\uparrow$ 
        & \textbf{Rank}$\downarrow$ \\
        \midrule
        VACE-Wan           & 8.3\%  & 3.95 & 2.33 & 1.92 & 5.58 \\
        VACE-LTX           & 8.3\%  & 3.51 & 1.58 & 1.50 & 6.00 \\
        Kiwi-Edit          & 16.7\% & 4.20 & 2.46 & 2.17 & 5.33 \\
        UniEdit            & 33.3\% & 2.88 & 3.12 & 2.08 & 7.33 \\
        LTX2.3 Retake      & 50.0\% & 5.77 & 4.67 & 4.29 & 3.00 \\
        Runway-Aleph       & 50.0\% & 5.56 & 5.46 & 4.88 & 3.58 \\
         LoRA-Edit          & 33.3\% & 5.37 & 4.33 & 3.88 & 3.58 \\
        \textbf{\paper{}}  & \textbf{83.3\%} & \textbf{7.16} & \textbf{7.42} & \textbf{6.48} & \textbf{1.58} \\
        \bottomrule
    \end{tabularx}
\end{table}

\subsection{Main Results}
\label{sec:exp_main_results}
 
\noindent\textbf{Qualitative Comparison.}
Fig.~\ref{fig:qualitative_ours_vs_all} compares \paper{} with representative baselines. Prompt-only Retake often preserves the source scene but misses the requested motion, while several general-purpose editors introduce visible changes at the cost of identity drift, layout changes, or abrupt object insertions. Our method more consistently realizes the requested temporal change while preserving the source subject and scene.

\noindent\textbf{Quantitative Comparison.}
Tables~\ref{tab:quantitative_human_survey} and~\ref{tab:quantitative_vlm} summarize human and VLM-based comparisons over the 12-scenario benchmark. In the human study, \paper{} receives 56.3\% of all first-place votes, more than twice the win rate of the strongest baseline, LoRA-Edit at 21.4\%. It also achieves the highest Top-3 Rate, AVG Score, and user-marked task achievement, indicating that participants prefer our outputs not only as the best result, but also consistently rank them among the strongest candidates.
GPT-5.5-Thinking evaluation confirms the same trend: \paper{} achieves the highest scores across all metrics and best average rank. Over LTX2.3 Retake, our method improves task success from 50.0\% to 83.3\% and motion quality from 4.29 to 6.48, and outperforms general-purpose editors in edit alignment and rank, confirming that test-time latent tuning is particularly effective for motion-centric edits.


\noindent\textbf{Runtime.}
 The run time for tuning the latents for our approach is roughly around 10 minutes. For instance, processing the turtle video (see Fig.\ref{fig:qualitative_ours_vs_all}) required approximately 589 seconds end-to-end. More details on per-iteration times and the hardware we used to run experiments are provided in supplementary materials Sec.~B.7.

\subsection{Ablation Studies}
\label{sec:exp_ablation}

\noindent\textbf{Effect of Joint Tuning.}
As shown in Fig.~\ref{fig:attention}, text and audio conditioning provide complementary controls: text-residual tuning can improve semantic alignment, while audio-latent tuning is often necessary for smooth and natural temporal motion. Fig.~\ref{fig:text_vs_both} further shows that in challenging scenarios, text-only tuning is insufficient, whereas joint tuning more reliably produces the desired edit. Quantitative ablations in Tab.~\ref{tab:ablation_modality} further support this observation: text tuning improves prompt semantic alignment, as reflected by the \emph{Alignment} score, while audio tuning improves motion naturalness and temporal coherence, as reflected by the \emph{Naturalness} score. Overall, joint tuning ranks first in \textbf{8 out of 10} scenarios (which includes a single tie in overall score in one scenario).

\noindent\textbf{Effect of Motion Supervision.}
Fig.~\ref{fig:ablation_scorer} compares different semantic supervision signals. CLIP-based guidance produces a nearly static yawning state: the mouth opens, but the result lacks a clear temporal transition because CLIP scores frames independently and does not handle video dynamics. XCLIP provides some temporal awareness, but its limited temporal window makes it less effective for longer actions and can introduce undesired global changes, such as shifts in lighting and the dog's appearance. In contrast, Qwen-based supervision better captures whether the requested action unfolds over time, leading to a more natural and temporally coherent edit.

\noindent\textbf{Effect of Regularization}
Fig.~\ref{fig:ablation_reg} shows the effect of regularization terms. Without regularization, VLM-guided tuning can introduce artifacts such as blur, inconsistent colors, and abrupt motion changes. Latent regularization constrains the search around the source-conditioned latents, reducing destructive drift. LPIPS preservation improves frame-level visual fidelity and color consistency, but does not fully prevent sudden temporal changes. Temporal regularization encourages smoother frame-to-frame evolution. Combining latent, LPIPS, and temporal regularization gives the best trade-off, producing cleaner frames and more coherent motion.

\subsection{Transferability of Optimized Latents}
\label{sec:exp_transfer}

To test generalization beyond a single clip, we optimize $\Phi_s^{*}=\{\mathbf{\alpha}^{*},\Delta\mathbf{v}^{*}\}$ on a source video with prompt $p_s$, then reuse the same controls on a different target video with a related prompt $p_t$ without further tuning, preserving the action while changing subject or scene, e.g., transferring ``a dog yawning'' to ``a cat yawning''. 

As shown in Fig.~\ref{fig:transferability_exp}, transferred latents can induce the corresponding motion in new videos while preserving target content. This suggests that the tuned audio-text controls capture reusable motion directions, rather than merely memorizing frame-level artifacts.

\section{Conclusions, Limitations and Future Work}
\label{sec:conclusion}

We presented a training-free approach for motion-centric video editing by tuning audio and residual text conditioning in a frozen audio-visual generator. This design also sheds light on a broader aspect of multimodal control. Strengthening multimodal conditioning does not necessarily translate to improved controllability. In such models, motion is not represented explicitly but instead emerges from complex interactions across modalities, and increased reliance on cross-modal attention can further entangle these interactions, making prompt-based control less reliable. At the same time, this structure reveals a complementary opportunity: the model already encodes latent motion controls within its conditioning pathways. By tuning these internal signals at test time, we can exploit these controls without modifying the model. This suggests a shift from introducing new controls to leveraging those already embedded in multimodal generative models.

Despite its effectiveness, our approach has two limitations. First, the test-time tuning process needs additional computation, which may limit interactivity. Second, the tuning does not always fully realize the desired motion, particularly for complex or ambiguous actions. 
In the absence of a general text-motion alignment model, our VLM-based binary feedback may be limited to coarse motions; a dedicated text-motion correspondence model would likely broaden the range of edits our method can support.

For future work, we plan to extend our approach beyond the current setting. Our current implementation targets the LTX model, as it is the only publicly available audio-visual video generation model with editing capabilities at this time, and we plan to evaluate our approach on additional models as they become available. In addition, reducing the cost of test-time tuning is an important direction. In particular, incorporating feedback earlier in the denoising process may allow for faster convergence and more efficient motion induction.

\newpage

\bibliographystyle{ACM-Reference-Format}
\bibliography{main}

\clearpage
\appendix
\begin{figure*}[t]
    \centering
    \ifshowsuppfigs
    \includegraphics[width=\textwidth]{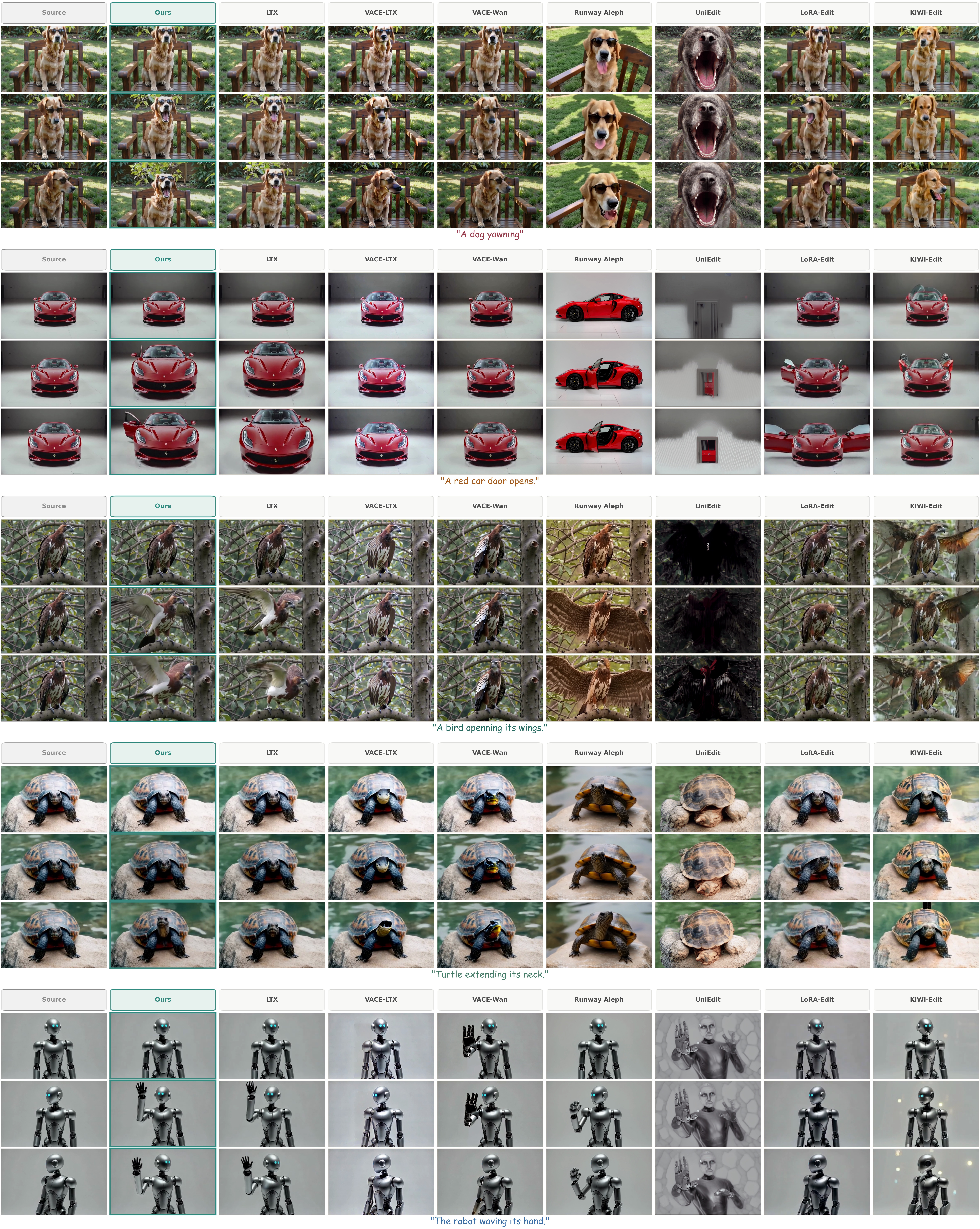}
    \fi
    \vspace{-2mm}
    \caption{\textbf{Qualitative Comparisons} of \paper{} vs various different editing methods on diverse set of edits.}
\label{fig:qualitative_ours_vs_all}
\end{figure*}

\vspace{-5mm}

\begin{figure*}[t]
    \centering
    \ifshowsuppfigs
    \includegraphics[width=\textwidth]{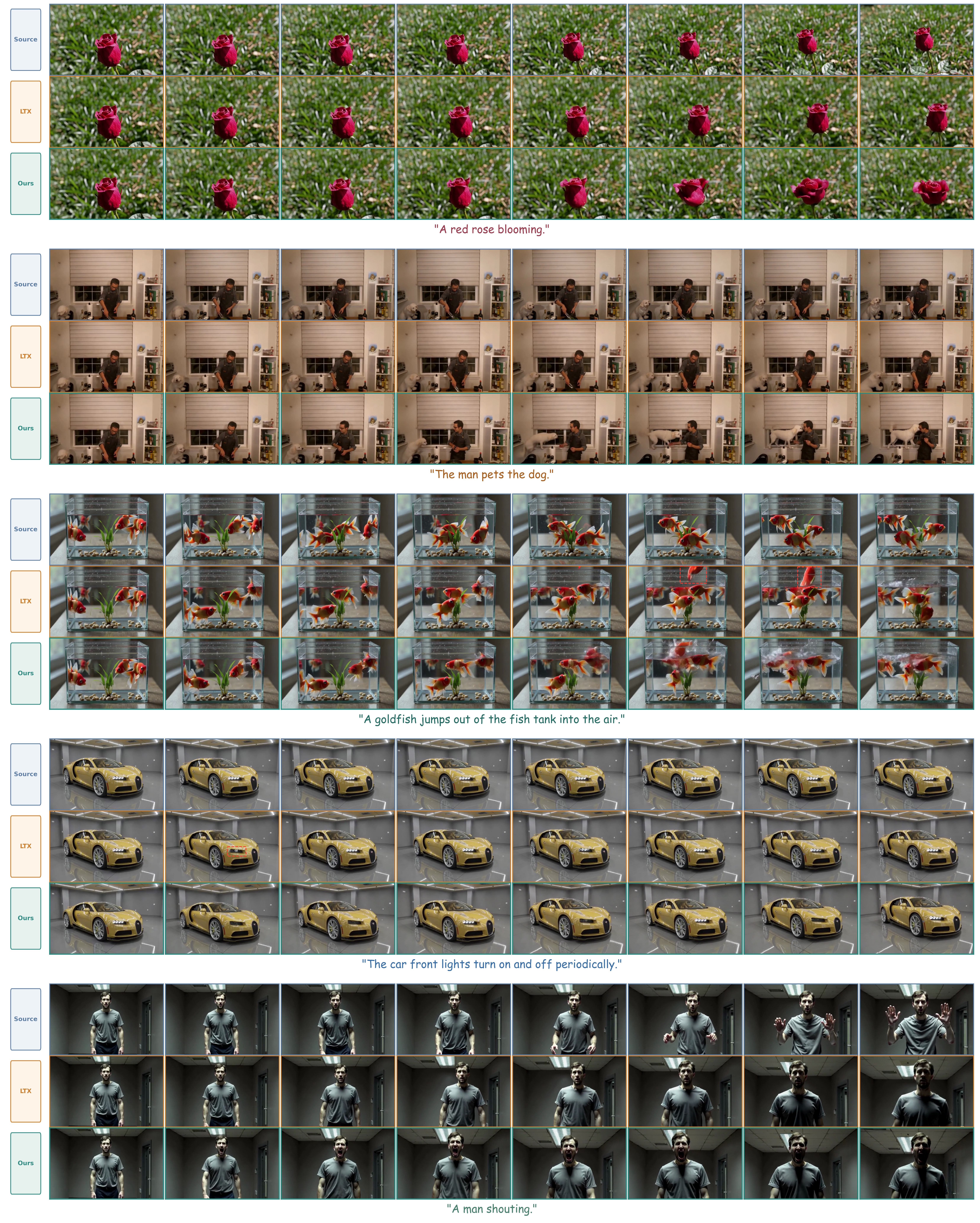}
    \fi
    \vspace{-2mm}
    \caption{\textbf{Qualitative Comparisons} of \paper{} vs LTX2.3 Retake on diverse set of edits. Red boxes show abrupt changes or magical insertions or obvious artifacts in the LTX Retake method.}
\label{fig:qualitative_ours_vs_ltx}
\end{figure*}

\makeatletter
\@ifundefined{@titlefont}
  {\newcommand{\supplementtitlefont}{\LARGE\bfseries}}
  {\newcommand{\supplementtitlefont}{\@titlefont}}
\makeatother

\clearpage
\twocolumn[
  \begin{center}
    {\supplementtitlefont Sound Sparks Motion: Audio and Text Tuning for Video Editing (Supplementary material)\par}
    \vspace{1.5em}
  \end{center}
]

\setcounter{figure}{0}
\renewcommand{\thefigure}{S\arabic{figure}}
\setcounter{table}{0}
\renewcommand{\thetable}{S\arabic{table}}
\setcounter{equation}{0}
\renewcommand{\theequation}{S\arabic{equation}}
\appendix

\section{LTX Retake Preliminaries}
\label{sec:supp_ltx_retake}

\subsection{LTX Audio--Video Generation}
\label{sec:supp_ltx_overview}

LTX2~\cite{ltx2} is a pretrained audio--video generative model that supports text, video, and audio conditioning. In this work, we use it only as a frozen editing backbone. At a high level, LTX2 maps input media and text into internal conditioning representations, runs a diffusion generator in latent space, and decodes the resulting latent representation back into video and audio.

The key property we rely on is its multimodal conditioning structure. The source video provides visual identity, layout, and scene context; the text prompt specifies the intended edit; and the audio pathway provides a temporally aligned conditioning signal coupled to the generated video. Although our goal is visual motion editing rather than audio editing, the audio-conditioning pathway gives us an additional internal handle over temporal dynamics. Our method exploits this handle without changing the pretrained LTX2 weights.

\subsection{Retake Conditioning and In-Context Regeneration}
\label{sec:supp_retake_conditioning}

LTX Retake is the editing mode of LTX2. Given a source video, a text prompt, and an editable temporal region, Retake preserves the reference part of the input and regenerates the selected region in context. In our setting, this means that the beginning of the source video is kept as conditioning, while the later frames are regenerated according to the edit prompt.

Conceptually, Retake builds three types of conditioning: visual conditioning from the source video, text conditioning from the prompt, and audio conditioning from the source audio. The source video anchors the scene and identity, the prompt describes the desired change, and the audio latent provides temporally structured information that is synchronized with the video generation process. A temporal mask determines which part of the video is preserved and which part is regenerated.

This retake formulation is a natural fit for motion-centric editing: the model can keep the initial subject and scene intact while synthesizing a new temporal continuation. However, prompt-only Retake often preserves the source well but fails to execute precise actions or state changes. This motivates our test-time tuning of the internal conditioning variables.

\subsection{How \paper{} Interfaces with Retake}
\label{sec:supp_retake_ours}

\paper{} does not finetune LTX2, train adapters, or rewrite the input prompt. Instead, it treats Retake as a frozen differentiable editor and optimizes only two lightweight conditioning variables at test time: the audio-conditioning latent $\mathbf{\alpha}$ and a residual text-conditioning vector $\Delta\mathbf{v}$.

We first run the standard Retake preprocessing to obtain the source video and audio representations. The source video conditioning is kept fixed to preserve the visual content. The source audio latent $\mathbf{\alpha}_0$ initializes the learnable audio latent $\mathbf{\alpha}$, and the frozen text encoder maps the edit prompt to a video text-conditioning representation $\mathbf{v}_p$. We then add a learnable residual $\Delta\mathbf{v}$ to this text representation:
\begin{equation}
    \mathbf{c}
    =
    \left(
        \mathbf{\alpha},
        \mathbf{v}_p + \Delta\mathbf{v}
    \right).
\end{equation}

During optimization, Retake is executed with the current conditioning state and produces an edited video. A frozen VLM critic evaluates whether the generated video satisfies the requested motion, and gradients update only $\mathbf{\alpha}$ and $\Delta\mathbf{v}$. All LTX2 components remain frozen. Thus, our method is best understood as test-time search in the internal multimodal conditioning space of Retake, rather than model finetuning or prompt engineering.


\begin{figure}[t]
    \centering
    \includegraphics[width=\linewidth]{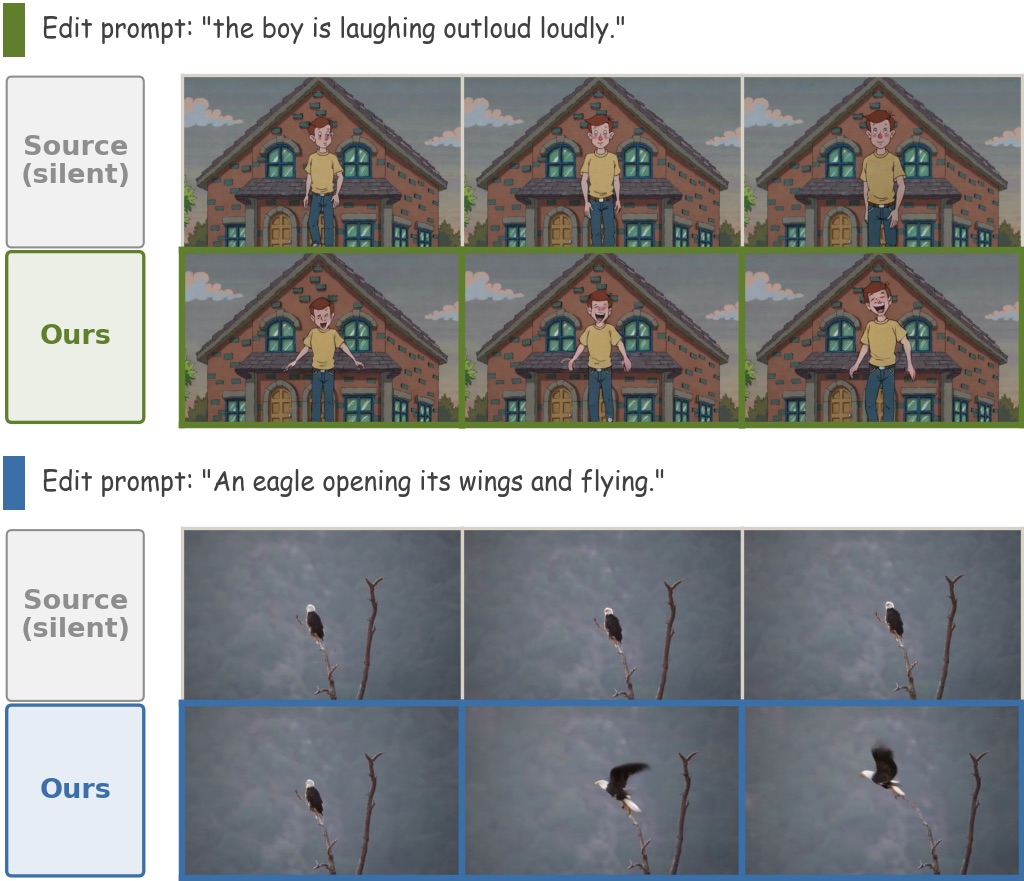}
    \vspace{-1.0em}
    \caption{
    \textbf{Editing videos without source audio.}
    For silent inputs, we first use frozen LTX Retake to synthesize audio while preserving all video frames. The generated audio initializes the audio-conditioning latent, after which \paper{} runs unchanged and tunes the audio and residual text latents for the requested visual motion edit.
    }
    \label{fig:supp_no_audio}
    \vspace{-1.0em}
\end{figure}

\section{Implementation Details}
\label{sec:supp_implementation}

Our method is implemented as a test-time tuning wrapper around the frozen LTX2 Retake pipeline. Unless otherwise stated, the LTX denoiser, video VAE, audio VAE, text encoder, audio encoder, and Qwen2.5-VL critic remain frozen throughout optimization.

\subsection{Input Preprocessing and Retake Configuration}
\label{sec:supp_input_retake_config}

Each input video is converted to the format expected by Retake. In our experiments, videos are edited at $512\times320$ resolution, 25 FPS, and 89 frames. The frame count satisfies the LTX temporal constraint $(T-1)\bmod 8=0$; if needed, preprocessing adjusts the requested length to satisfy this constraint.

We configure Retake to preserve an initial prefix of the source video and regenerate the remaining suffix in context. Let $K$ denote the number of preserved conditioning frames. The retake start time is
\begin{equation}
    t_{\mathrm{start}} = K / f,
\end{equation}
where $f$ is the video frame rate. The retake end time is set to the full clip duration. In our experiments, $K$ varies from 3 to 25 depending on edit complexity and the desired strength of source preservation.

We use \texttt{regenerate\_video=True} and \texttt{regenerate\_audio=False}. Therefore, the video suffix is regenerated, while audio is not decoded as an edited output during tuning. The audio latent is used only as an internal conditioning variable for controlling the frozen audio--video generator.

\subsection{Latent Parameterization and Initialization}
\label{sec:supp_latent_parameterization}

We cache the source video and audio representations once before optimization. The source video latent remains fixed throughout the edit. The source audio is encoded into an audio-conditioning latent $\mathbf{\alpha}_0$, which initializes the learnable audio latent $\mathbf{\alpha}$.

The edit prompt is encoded once by the frozen text encoder. Let $\mathbf{v}_p$ denote the video-conditioning component of the prompt representation. We optimize a residual text-conditioning vector $\Delta\mathbf{v}$, initialized to zero, and use
\begin{equation}
    \tilde{\mathbf{v}}_p = \mathbf{v}_p + \Delta\mathbf{v}
\end{equation}
as the effective text conditioning during tuning. The prompt tokens are kept fixed, and no prompt rewriting is performed.

The only optimized variables are
\begin{equation}
    \Phi =
    \left\{
        \mathbf{\alpha},
        \Delta\mathbf{v}
    \right\},
    \qquad
    \mathbf{\alpha}^{(0)} = \mathbf{\alpha}_0,
    \qquad
    \Delta\mathbf{v}^{(0)} = \mathbf{0}.
\end{equation}
Thus, tuning is performed over internal conditioning latents, not over model weights, pixels, diffusion noise, or prompt tokens.

\subsection{Differentiable Test-Time Execution}
\label{sec:supp_differentiable_execution}

The original Retake pipeline is designed for inference, so we wrap it to expose gradients with respect to $\mathbf{\alpha}$ and $\Delta\mathbf{v}$. At each optimization step, Retake receives the fixed source video conditioning, the current audio latent $\mathbf{\alpha}$, and the current text context $\mathbf{v}_p+\Delta\mathbf{v}$. The decoded video frames are returned as differentiable tensors so that the VLM and preservation losses can update the optimized latents.

Backpropagating through the entire diffusion trajectory is memory-intensive. Therefore, we propagate gradients only through the final denoising steps and keep earlier steps fixed during backpropagation. In our final configuration, we use the last 8 denoising steps for gradient propagation. This affects only the tuning procedure; the pretrained model weights and final inference pipeline are not modified.

\subsection{Qwen Supervision and Temporal Sampling}
\label{sec:supp_qwen_sampling}

We use Qwen2.5-VL-7B-Instruct~\cite{qwen} as a frozen video-language critic. For each generated video, we ask a binary motion question:
\begin{quote}
Does this video clearly show the action or state change described by the edit prompt?
\end{quote}
The logits assigned to the answer tokens \texttt{yes} and \texttt{no} are extracted at the final response position. The yes probability is computed as a binary softmax
\begin{equation}
    P_{\mathrm{yes}}
    =
    \frac{
        \exp(\ell_{\mathrm{yes}})
    }{
        \exp(\ell_{\mathrm{yes}})
        +
        \exp(\ell_{\mathrm{no}})
    },
\end{equation}
and the VLM loss is
\begin{equation}
    \mathcal{L}_{\mathrm{vlm}}
    =
    -\log(P_{\mathrm{yes}}+\epsilon),
\end{equation}
where $\epsilon$ is a small constant for numerical stability.

\begin{figure*}[t]
    \centering
    \includegraphics[width=\textwidth]{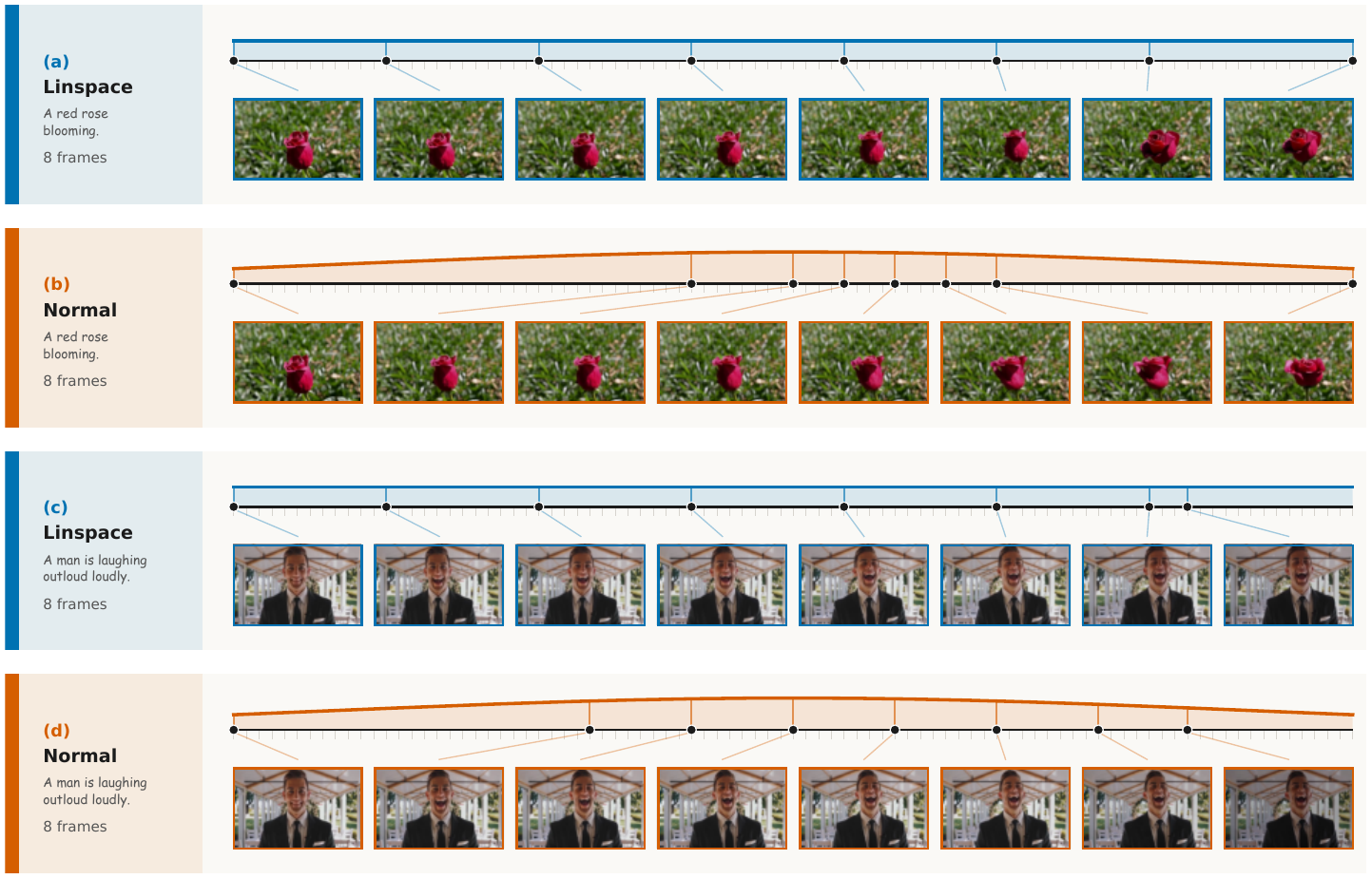}
    \vspace{-1.5em}
    \caption{
    \textbf{Effect of temporal sampling for VLM supervision.}
    Our method succeeds under different frame-sampling schedules for Qwen-based supervision. We compare uniform sampling (\emph{Linspace}) and midpoint-biased sampling (\emph{Normal}) using 8 VLM input frames. While both schedules induce the requested motion, the sampling strategy can optionally serve as a lightweight temporal control handle: uniform sampling distributes supervision across the full clip, whereas midpoint-biased sampling emphasizes the central editable region and can slightly affect the timing or progression of the generated motion.
    }
    \label{fig:supp_frame_sampling}
    \vspace{-1.0em}
\end{figure*}

For efficiency, Qwen is applied to a sampled subset of frames rather than the full video. We use either \textbf{uniform sampling}, which spreads supervision across the clip, or \textbf{midpoint-biased sampling}, which focuses more strongly on the central editable region. As shown in Fig.~\ref{fig:supp_frame_sampling}, both strategies can produce successful motion edits, while the sampling schedule can optionally provide a lightweight control over the timing and progression of the generated motion.

\begin{figure}[t]
    \centering
    \includegraphics[width=\linewidth]{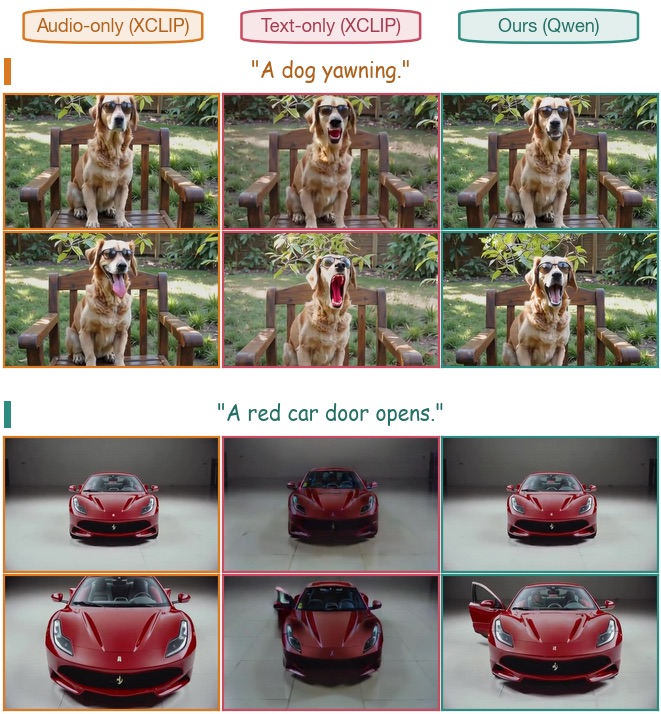}
    \vspace{-1.0em}
    \caption{
    \textbf{Extended visualization of latent and supervision effects.}
    Text-only tuning with XCLIP supervision tends to alter global appearance, such as lighting and color, and can produce exaggerated unnatural motion. Audio-only tuning better preserves source consistency, but the induced motion is often too subtle. Joint audio--text tuning with Qwen supervision achieves a better balance, producing clearer, smoother, and more realistic motion while preserving the source video.
    }
    \label{fig:supp_text_audio_xclip}
    \vspace{-1.0em}
\end{figure}

\subsection{Regularization and Preservation Losses}
\label{sec:supp_regularization}

We regularize both optimized variables to prevent latent drift:
\begin{equation}
    \mathcal{L}_{\mathrm{latent}}
    =
    \lambda_{\alpha}
    \|\mathbf{\alpha}-\mathbf{\alpha}_0\|_2^2
    +
    \lambda_v
    \|\Delta\mathbf{v}\|_2^2 .
\end{equation}
We use $\lambda_{\alpha}=0.01$ and $\lambda_v=0.001$ in our experiments.

We also use perceptual and temporal preservation losses when enabled. The LPIPS preservation term compares the optimized output to the prompt-only Retake baseline, which anchors the edit to a clean Retake result while still allowing the requested motion to emerge. The temporal preservation term penalizes excessive frame-to-frame perceptual variation relative to the baseline, reducing flicker and abrupt changes without suppressing natural motion.

The qualitative and quantitative ablations in Fig.~\ref{fig:supp_ablation_reg} and Fig.~\ref{fig:supp_reg_ablation} show that these terms are complementary: latent regularization reduces drift, LPIPS improves visual fidelity, and temporal regularization improves motion smoothness.

\subsection{Optimization Hyperparameters}
\label{sec:supp_optimization_hyperparams}

We optimize $\mathbf{\alpha}$ and $\Delta\mathbf{v}$ jointly using Adam~\cite{Adam}. The final configuration uses 30 optimization iterations, early stopping patience of 15 iterations, learning rate $5\times10^{-3}$, cosine learning-rate decay, and no gradient clipping. The final optimized video is rendered using the best checkpoint selected according to the total objective.

During optimization, we optionally use lower-memory guidance settings for efficiency. After optimization, the final video is rendered again with the optimized latents and the full-quality/default guidance configuration.

\subsection{Runtime, Memory, and Hardware}
\label{sec:supp_runtime}

All experiments are run on a single NVIDIA H100 80GB GPU. We use the LTX2.3 22B checkpoint with FP8-cast quantization during optimization, together with gradient checkpointing and low-memory guidance. Qwen2.5-VL-7B-Instruct is used as the frozen VLM critic, and Gemma-3-12B~\cite{gemma3} as the text encoder. Videos are optimized at $512\times320$ resolution with 30 Retake denoising steps.

Runtime scales roughly linearly with the number of test-time tuning iterations. In a representative 15-iteration run without early stopping, tuning took $589.6$ seconds: the warmup iteration took $41.26$ seconds, and later iterations averaged $38.87$ seconds ($38.34$--$40.04$). Our default setting uses up to 30 iterations with early stopping patience 15, followed by one final render from the selected best checkpoint. All representative runs fit on a single H100 80GB GPU.


\section{Editing Videos Without Source Audio}
\label{sec:supp_no_audio}

Our method uses the audio-conditioning pathway as an internal temporal control variable. When an input video does not contain an audio track, we first synthesize a compatible audio signal using the frozen LTX Retake pipeline, and then run our standard latent-tuning procedure.

Specifically, we pass the source video to Retake pipeline with \texttt{regenerate\_video=False} and \texttt{regenerate\_audio=True}. Equivalently, the reference mask covers the full video: all video frames are preserved, and only the audio stream is generated. This produces an audio track conditioned on the visual content of the input video. We then combine this generated audio with the original source video and use the resulting video-audio pair as input to \paper{}.

The rest of the pipeline is unchanged. The generated audio is encoded into the initial audio-conditioning latent $\mathbf{\alpha}_0$, and we optimize $\mathbf{\alpha}$ and $\Delta\mathbf{v}$ using the same VLM-guided objective. Fig.~\ref{fig:supp_no_audio} shows an example of this procedure. Importantly, the generated audio is used only to initialize the internal audio-conditioning pathway; the optimization still targets the requested visual edit.

\begin{figure}[t]
    \centering
    \includegraphics[width=\linewidth]{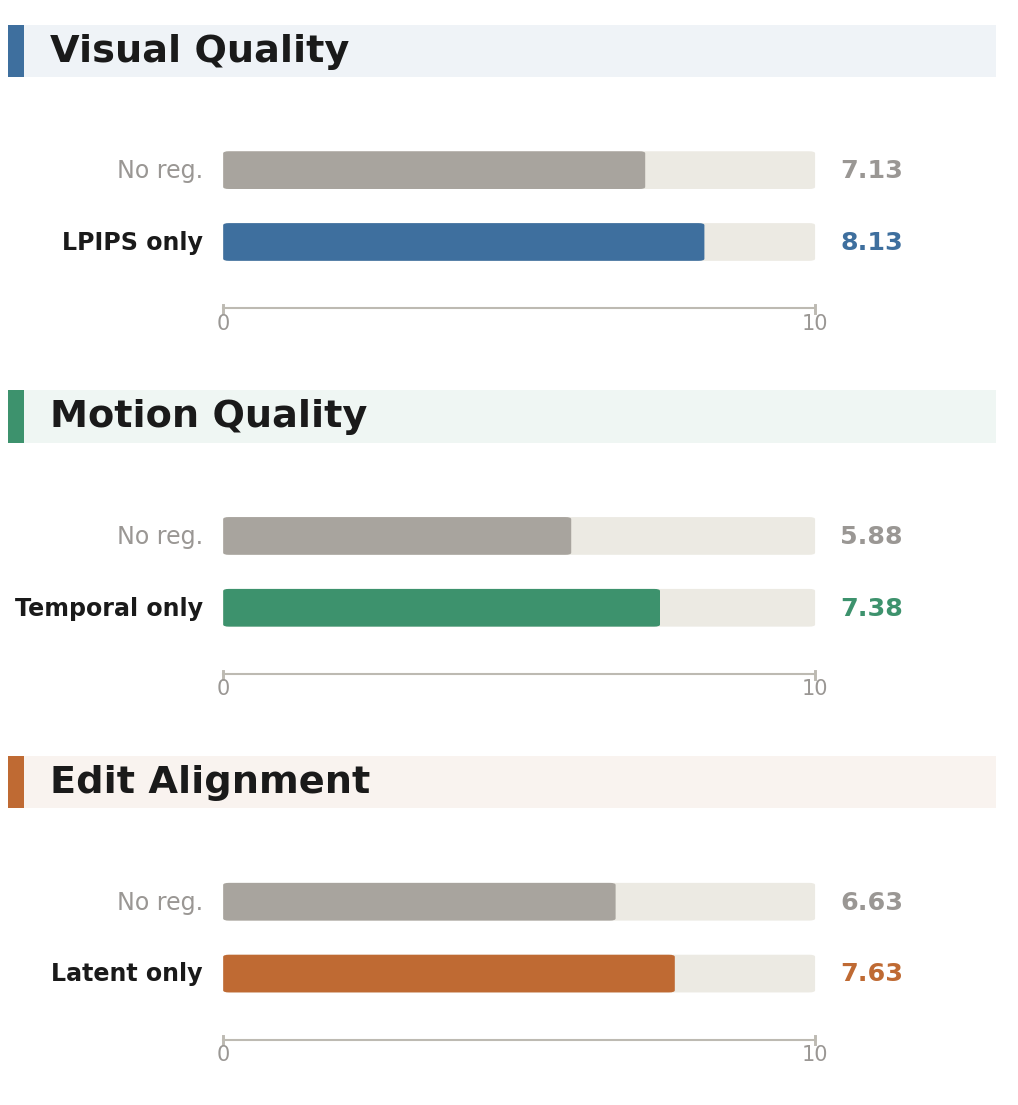}
    \vspace{-1.0em}
    \caption{
    \textbf{Quantitative effect of regularization terms.}
    We report 0--10 quality scores, where \textbf{higher is better}. Compared to tuning without regularization, LPIPS preservation improves visual quality, temporal regularization improves motion quality, and latent regularization improves edit alignment by reducing destructive latent drift. These results support using the full regularized objective, where the terms act as complementary safeguards against artifacts, abrupt motion, and unintended scene changes.
    }
    \label{fig:supp_reg_ablation}
    \vspace{-1.0em}
\end{figure}

\begin{figure}[t]
    \centering
    \includegraphics[width=\linewidth]{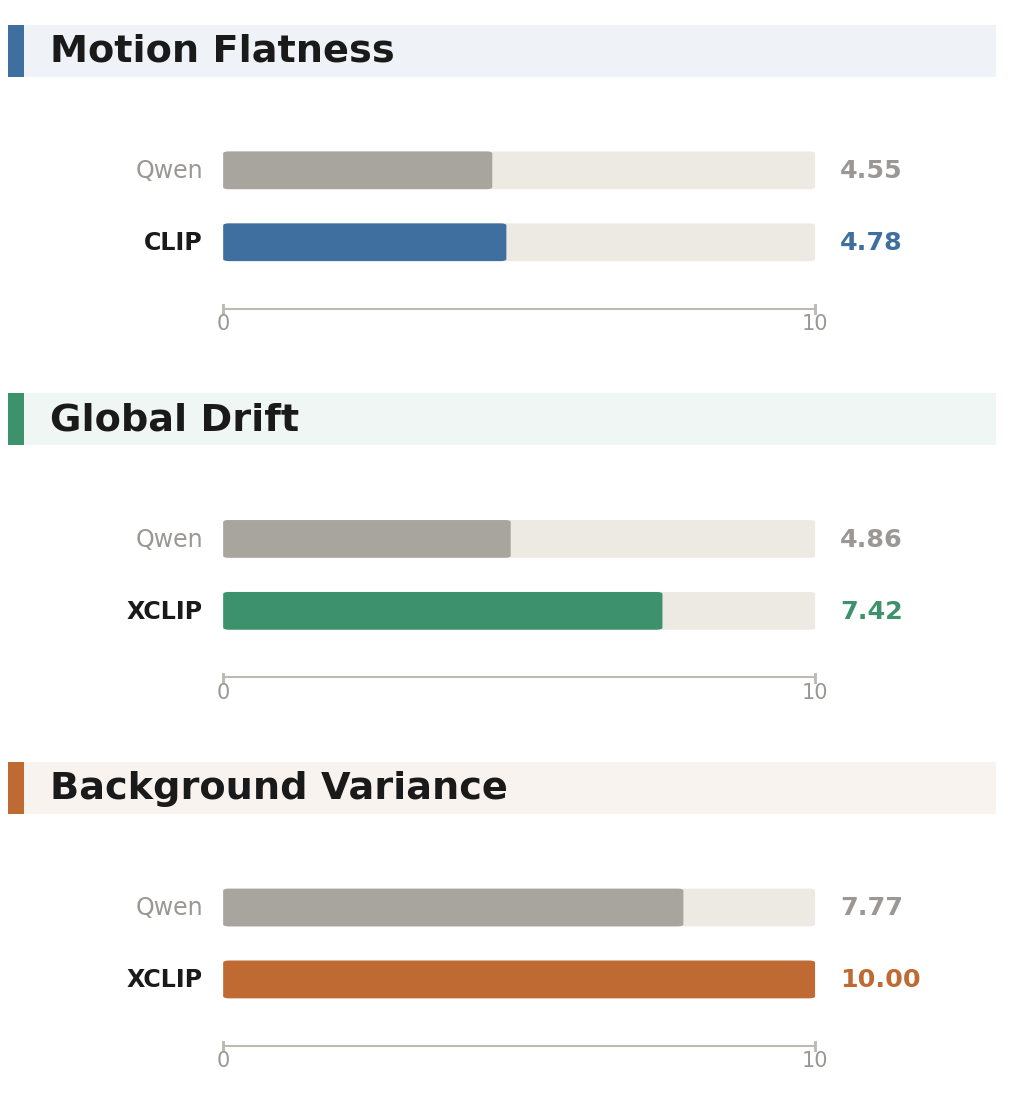}
    \vspace{-1.0em}
    \caption{
    \textbf{Quantitative effect of semantic supervision.}
    We compare supervision methods using motion flatness, global drift, and background variance; \textbf{lower is better} for all three metrics. Compared to CLIP and XCLIP guidance, Qwen supervision produces lower scores, indicating stronger temporal edit behavior with fewer unintended scene changes. These results support using a vision-language critic that can judge the requested action over time rather than relying on frame-level or short-window similarity scores.
    }
    \label{fig:supp_scorer_ablation}
    \vspace{-1.0em}
\end{figure}


\section{Quantitative Ablation Results}
\label{sec:supp_quant_ablation}

Fig.~\ref{fig:supp_reg_ablation} reports quantitative scores for the regularization ablation. We evaluate each variant using the same 0--10 VLM-based criteria used in the main evaluation: edit alignment, motion quality, source preservation, and visual quality. These scores are computed from the generated video, source video, and edit prompt, and higher values indicate better performance. The results support the qualitative observations in Fig.~\ref{fig:supp_ablation_reg}: each regularization term improves a different aspect of the edit, while the full objective provides the best overall balance between edit strength, visual fidelity, and temporal stability.

\begin{figure*}[t]
    \centering
    \includegraphics[width=0.9\textwidth]{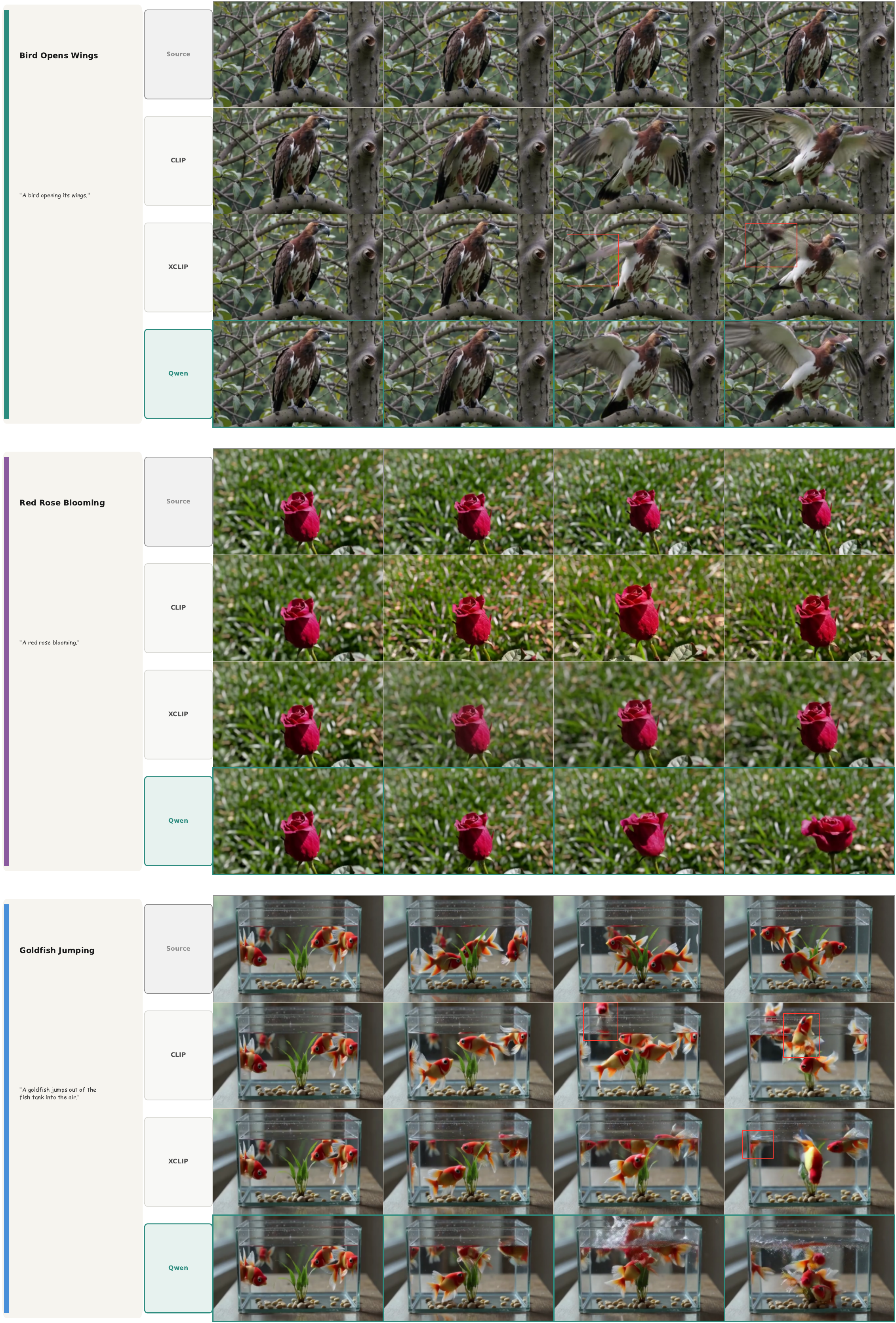}
    \vspace{-2mm}
    \caption{
    \textbf{Additional supervision ablations.}
    CLIP-style frame-level guidance often encourages static semantic appearance, while XCLIP provides limited temporal awareness. Qwen-based motion supervision better captures whether the requested action unfolds over time.
    }
    \label{fig:supp_ablation_scorer}
\end{figure*}

Fig.~\ref{fig:supp_scorer_ablation} reports quantitative metrics for the semantic-supervision ablation. We measure three failure modes that are visible in Fig.~\ref{fig:supp_ablation_scorer}. First, \emph{motion flatness} is computed from dense optical flow after removing the dominant global translation; it measures whether local motion remains overly constant or static-like rather than forming a clear temporal action. Second, \emph{global drift} measures the dominant frame-level optical-flow translation, normalized per scenario, and captures camera/layout shifts. Third, \emph{background variance} measures unintended background changes by first estimating foreground masks with SAM2~\cite{sam2} and then comparing source and edited videos only on background pixels using LPIPS~\cite{lpips}, MS-SSIM~\cite{wang2003multiscale}, and $\ell_1$ differences:
\begin{equation}
    D_{\mathrm{bg}}
    =
    0.5\,D_{\mathrm{LPIPS}}
    +
    0.3\,(1-D_{\mathrm{MS\text{-}SSIM}})
    +
    0.2\,D_{\ell_1}.
\end{equation}
The raw background-change value is normalized within each scenario to a 0--10 score. For all three metrics in Fig.~\ref{fig:supp_scorer_ablation}, lower is better. CLIP~\cite{clip} and XCLIP~\cite{xclip} provide useful semantic signals, but are less reliable for judging whether the requested action unfolds over time. Qwen supervision yields lower motion flatness, global drift, and background variance, matching the qualitative examples in Fig.~\ref{fig:supp_ablation_scorer}.

\begin{figure}[t]
    \centering
    \includegraphics[width=\linewidth]{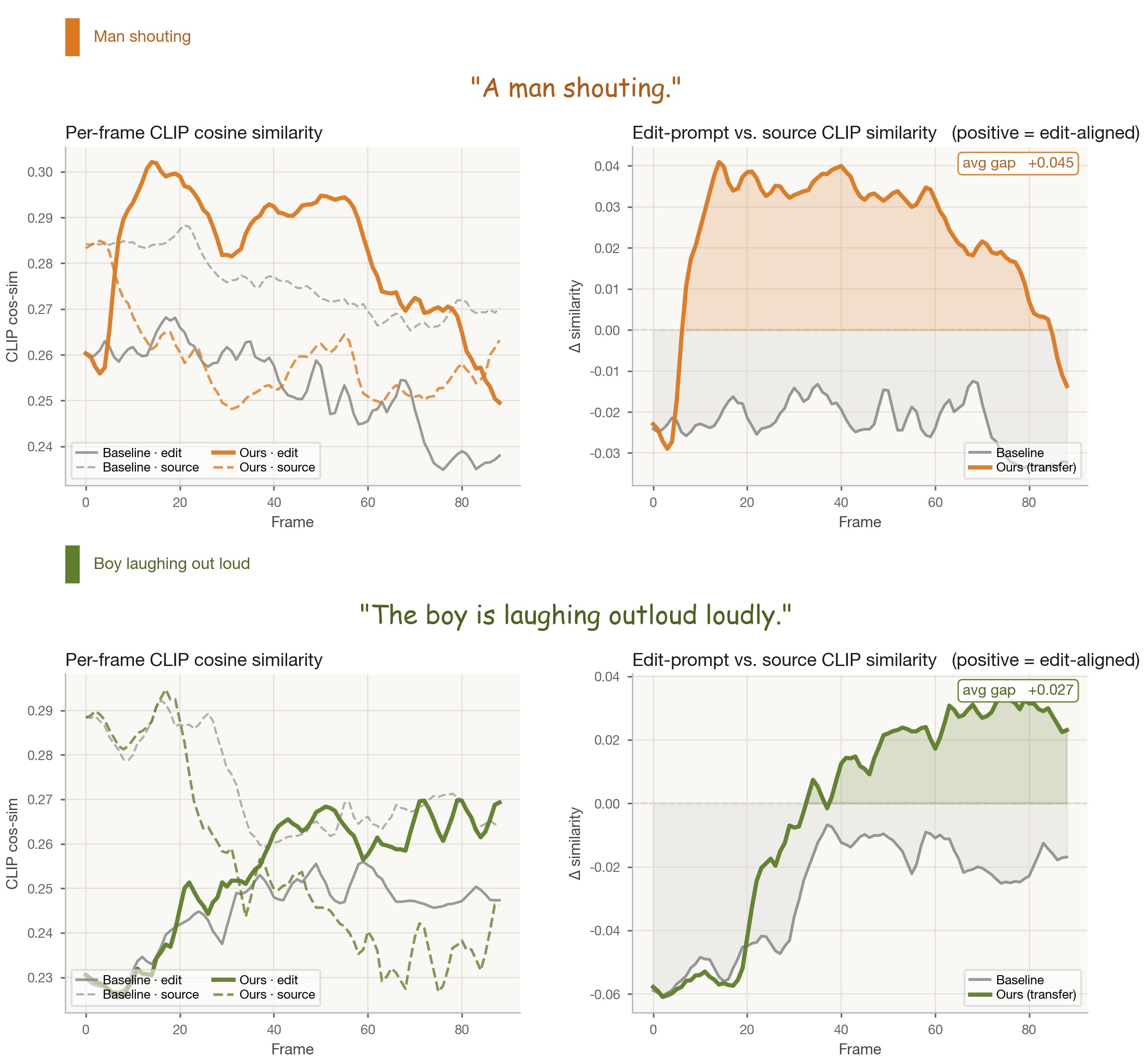}
    \vspace{-1.0em}
    \caption{
    \textbf{Quantitative evidence for latent transferability.}
    Transferred latents increase per-frame CLIP similarity to the target edit prompt while reducing similarity to the source prompt. The positive edit--source similarity gap shows that the reused latents shift new videos toward the desired action more strongly than the LTX baseline.
    }
    \label{fig:supp_transferability_quantitative}
    \vspace{-1.0em}
\end{figure}

\section{Extended Quantitative Evaluation}
\label{sec:supp_extended_quant}

Table~\ref{tab:human_survey_per_scenario} provides the per-scenario breakdown of the human evaluation results. In the main paper, we report aggregate results over all evaluated scenarios; here, we include detailed scores for each individual task to show how human preferences vary across edit types.

The per-scenario results show that \paper{} receives the highest human preference in most motion-centric tasks, including car lights, goldfish jumping, man climbing, man petting dog, man raising hand, dog yawning, monkey reaching fruit, robot waving, and turtle neck extension. The table also highlights competitive cases. In particular, LoRA-Edit performs strongly on rose blooming and red car door opening. This result should be interpreted with its input setting in mind: unlike \paper{}, which uses only the source video and text edit prompt at test time, LoRA-Edit is additionally supervised with edited reference frames. In our comparison, we provide LoRA-Edit with GPT-Imagegen2 generated edited first and last frames, which guide both the start and end states of the edit. Thus, LoRA-Edit represents a strong reference-frame-supervised baseline rather than a purely prompt-driven editor. Despite this additional supervision, \paper{} obtains stronger human preference across most evaluated motion edits.

Table~\ref{tab:supp_quantitative_vlm} extends the automatic VLM evaluation with source preservation (SP) and visual quality (VQ) scores alongside the motion-focused metrics reported in the main paper. A key observation is that SP must be interpreted jointly with task success: a method that performs no edit at all would achieve perfect source preservation by definition. This is reflected in the results---LTX2.3 Retake achieves the highest SP (7.92) and VQ (7.54) among all methods, yet succeeds on only 50\% of tasks (TS), indicating that in many cases it reproduces the source content without realizing the requested motion. Similarly, methods such as VACE-Wan and Kiwi-Edit obtain moderate SP scores (6.79 and 7.35, respectively) while achieving near-zero task success (8.3\% and 16.7\%), suggesting that their preservation comes at the cost of failing to edit. UniEdit represents the opposite failure mode: low SP (2.67) and VQ (3.67) indicate identity collapse without meaningful motion benefit.

\paper{} achieves the second-highest SP (7.49) and VQ (7.23) \emph{among methods that succeed on the task}, with a task success rate of 83.3\%. The modest SP gap relative to LTX2.3 Retake (7.49 vs.\ 7.92) is expected: inducing new motion---such as a dog opening its mouth, a turtle extending its neck, or a car door swinging open---necessarily involves localized changes to subject pose or form that a purely preservation-oriented metric will penalize. Taken together, these results indicate that \paper{} achieves the best balance between motion realization and source fidelity across all evaluated methods.


\begin{table}[t]
    \centering
    \setlength{\tabcolsep}{3.0pt}
    \renewcommand{\arraystretch}{1.08}
    \caption{
    \textbf{Extended metrics for GPT-based VLM evaluation across 12 scenarios.}
    TS is task success rate; VLM Score is the weighted 0--10 score; EA and MQ denote edit alignment and motion quality; SP and VQ denote source preservation and visual quality. Rank is the average comparative rank assigned by the VLM, lower is better. \textbf{Bold} indicates the best result per column; \underline{underline} indicates the second best.
    }
    \label{tab:supp_quantitative_vlm}
    \begin{tabularx}{\columnwidth}{@{}Xlcccccc@{}}
        \toprule
        \textbf{Method} 
        & \textbf{TS}$\uparrow$ 
        & \textbf{VLM}$\uparrow$ 
        & \textbf{EA}$\uparrow$ 
        & \textbf{MQ}$\uparrow$ 
        & \textbf{SP}$\uparrow$
        & \textbf{VQ}$\uparrow$
        & \textbf{Rank}$\downarrow$ \\
        \midrule
        VACE-Wan           & 8.3\%  & 3.95 & 2.33 & 1.92 & 6.79 & 6.46 & 5.58 \\
        VACE-LTX           & 8.3\%  & 3.51 & 1.58 & 1.50 & 6.50 & 6.42 & 6.00 \\
        Kiwi-Edit          & 16.7\% & 4.20 & 2.46 & 2.17 & 7.35 & 6.67 & 5.33 \\
        UniEdit            & 33.3\% & 2.88 & 3.12 & 2.08 & 2.67 & 3.67 & 7.33 \\
        LTX2.3 Retake      & \underline{50.0}\% & \underline{5.77} & 4.67 & 4.29 & \textbf{7.92} & \textbf{7.54} & \underline{3.00} \\
        Runway-Aleph       & \underline{50.0}\% & 5.56 & \underline{5.46} & \underline{4.88} & 5.08 & 7.08 & 3.58 \\
         LoRA-Edit          & 33.3\% & 5.37 & 4.33 & 3.88 & 7.38 & 7.07 & 3.58 \\
        \textbf{Ours}  & \textbf{83.3\%} & \textbf{7.16} & \textbf{7.42} & \textbf{6.48} & \underline{7.49} & \underline{7.23} & \textbf{1.58} \\
        \bottomrule
    \end{tabularx}
\end{table}

\begin{figure*}[t]
    \centering
    \includegraphics[width=\linewidth]{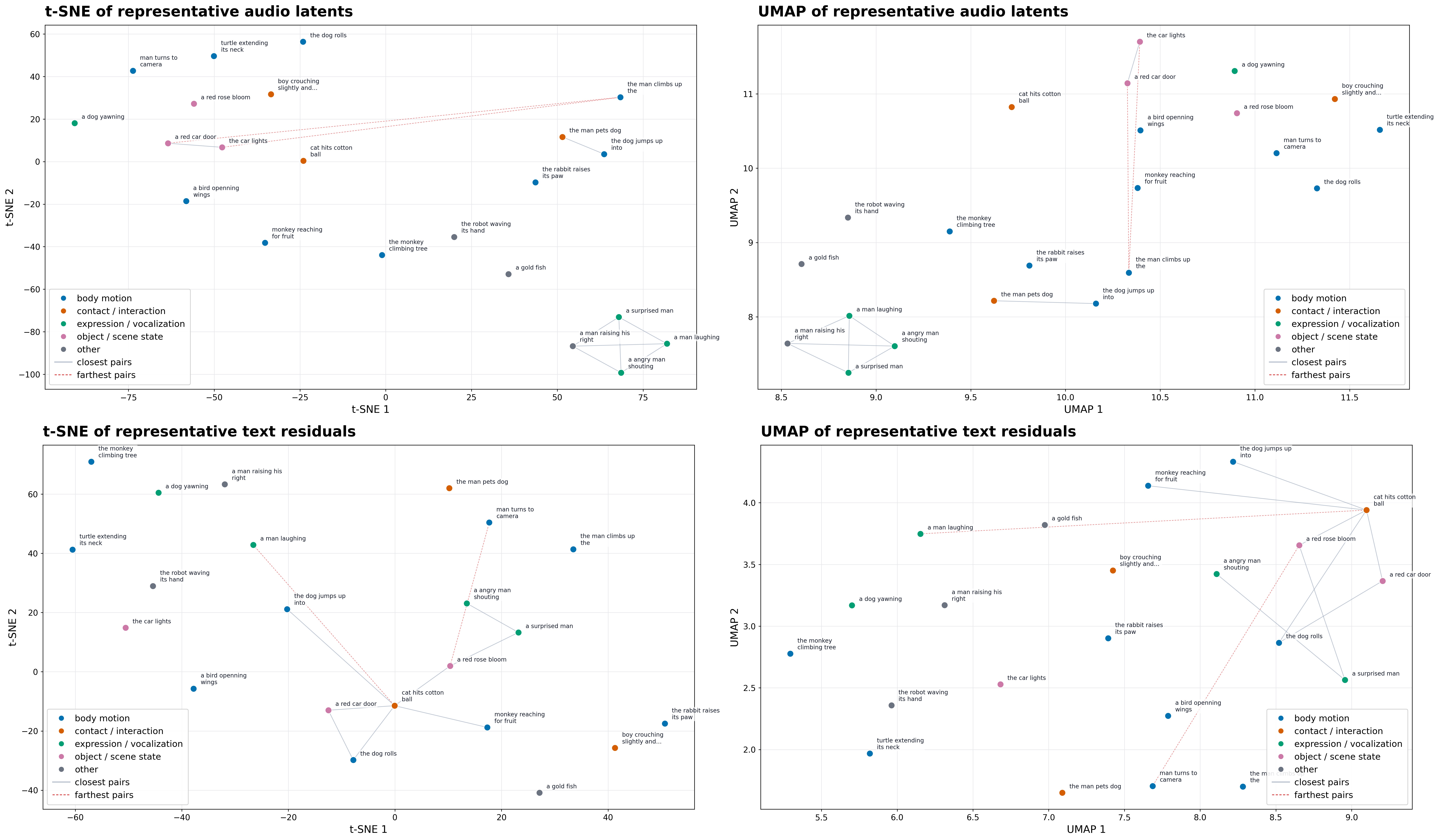}
    \vspace{-1.2em}
    \caption{
    \textbf{Diagnostic visualization of tuned audio latents and text residuals.}
    We visualize one representative best-run sample per prompt using t-SNE and UMAP. The top row shows tuned audio latents, while the bottom row shows tuned text residuals. Points are colored by manually assigned edit type; gray edges indicate nearest pairs in the original cosine space and red dashed edges indicate farthest pairs. The projections suggest that audio latents are more organized by implied motion dynamics, while text residuals preserve more object-level semantic similarity.
    }
    \label{fig:supp_audio_text_tsne_vis}
    \vspace{-1.2em}
\end{figure*}

\section{Additional Ablation Results}
\label{sec:supp_ablation}

\subsection{Additional Regularization Ablations}
\label{sec:supp_ablation_regularization}

\begin{figure*}[t]
    \centering
    \includegraphics[width=\textwidth]{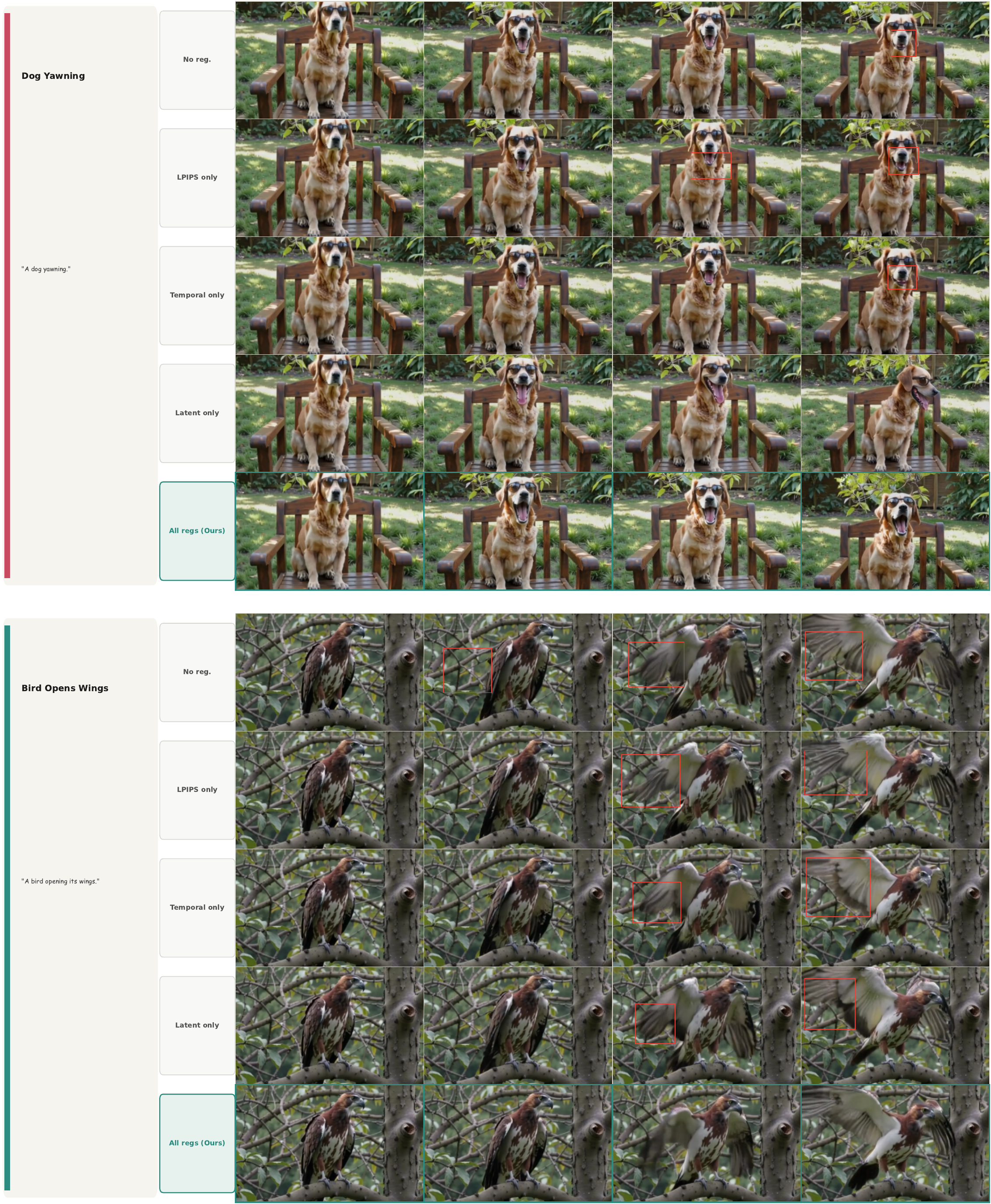}
    \vspace{-2mm}
    \caption{
    \textbf{Additional regularization ablations.}
    We show more examples comparing no regularization, latent-only, LPIPS-only, temporal-only, and the full objective. Combining all terms best reduces visual drift, flicker, and abrupt motion changes.
    }
    \label{fig:supp_ablation_reg}
\end{figure*}

Fig.~\ref{fig:supp_ablation_reg} provides additional qualitative examples for the regularization ablation. Without regularization, VLM-guided tuning can overfit to the critic and introduce artifacts, abrupt motion, or visual drift. Latent, LPIPS, and temporal regularization address complementary failure modes, and the full objective produces the most stable edits across examples.

\subsection{Additional Supervision Ablations}
\label{sec:supp_ablation_supervision}

Fig.~\ref{fig:supp_ablation_scorer} provides additional qualitative examples comparing CLIP, XCLIP, and Qwen supervision. CLIP often encourages frame-level semantic appearance rather than temporal action. XCLIP provides some temporal awareness, but its short temporal window can still miss longer motion structure. Qwen-based binary motion supervision better captures whether the requested action or state change unfolds over time.

\subsection{Additional Audio--Text Latent Visualizations}
\label{sec:supp_ablation_text_vs_audio}

We further inspect the optimized representations by projecting representative best-run audio latents and text residuals with t-SNE~\cite{van2008visualizing} and UMAP~\cite{mcinnes2018umap} in Fig.~\ref{fig:supp_audio_text_tsne_vis}. The nearest-pair structure suggests different tendencies in the two optimized spaces. In the text-residual space, prompts involving related agents or objects, such as \textit{cat hits the cotton ball}, \textit{dog jumps up}, and \textit{monkey reaching for a fruit}, tend to remain relatively close, suggesting that text residuals preserve object- and actor-level semantic structure. In contrast, the corresponding audio latents are more separated, consistent with the different temporal dynamics implied by hitting, jumping, and reaching. Conversely, prompts with different semantics but related motion structure can become closer in the audio space; for example, upward-motion edits such as \textit{dog jumps up} and \textit{man climbs up} appear nearby. Overall, this visualization supports our interpretation that audio latents provide a stronger handle on motion dynamics, while text residuals retain more semantic information about the entities and objects involved in the edit.

\begin{figure*}[t]
    \centering
    \includegraphics[width=\linewidth]{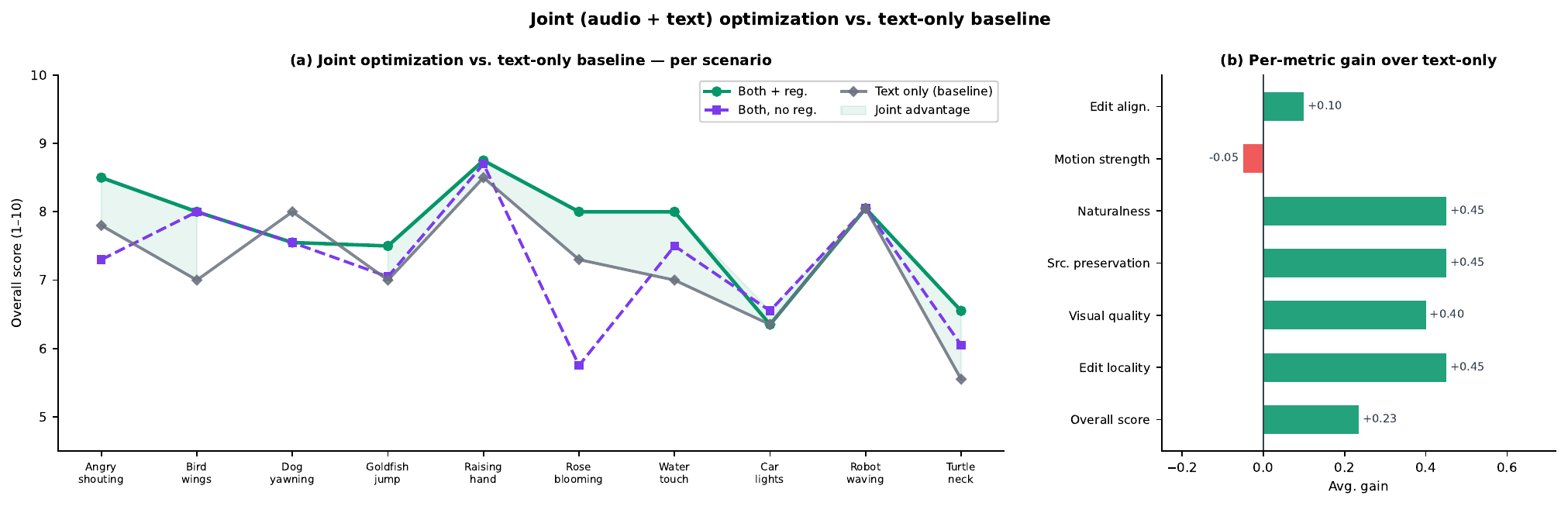}
    \vspace{-1.2em}
    \caption{
    \textbf{Joint audio--text tuning vs. text-only residual tuning.}
    Extended view of the modality ablation in Tab.~1 of main paper. 
    (a) Per-scenario overall scores for text-only tuning, unregularized joint tuning, and regularized joint tuning. 
    (b) Average metric-wise gain of regularized joint tuning over text-only tuning according to the VLM evaluator. 
    Joint tuning improves naturalness, preservation, visual quality, and edit locality, suggesting that the audio latent provides a temporal control beyond the text residual alone.
    }
    \label{fig:supp_text_vs_joint}
    \vspace{-1.2em}
\end{figure*}

\subsection{Direct Latent Optimization vs. PPO}
\label{sec:supp_ppo_ablation}

A natural alternative to our differentiable test-time tuning is to treat video editing as a black-box reinforcement-learning problem. PPO-style updates~\cite{ppo} are useful when gradients through the target system are unavailable or unreliable. In our setting, however, the editable variables are continuous internal latents and our wrapped Retake pipeline provides gradients from the VLM objective back to these variables. We therefore compare our gradient-based optimizer with a PPO baseline under matched conditions.

Both methods edit the same variables: the audio-conditioning latent $\mathbf{\alpha}$ and the residual text-conditioning vector $\Delta\mathbf{v}$. Both use the same frozen Retake renderer, source video, edit prompt, Qwen2.5-VL motion objective, regularizers, and final rendering path. The only difference is the update rule.

In the PPO baseline, a per-example Gaussian policy samples complete audio/text latent updates. Each sampled action renders a full candidate video and receives reward
\begin{equation}
    r =
    -
    \left(
        \mathcal{L}_{\mathrm{vlm}}
        +
        \mathcal{L}_{\mathrm{latent}}
        +
        \mathcal{L}_{\mathrm{preserve}}
    \right).
\end{equation}
Thus, PPO is not given an easier objective; it solves the same test-time editing problem using scalar rewards from full-video rollouts.

\begin{figure*}[t]
    \centering
    \includegraphics[width=0.75\textwidth]{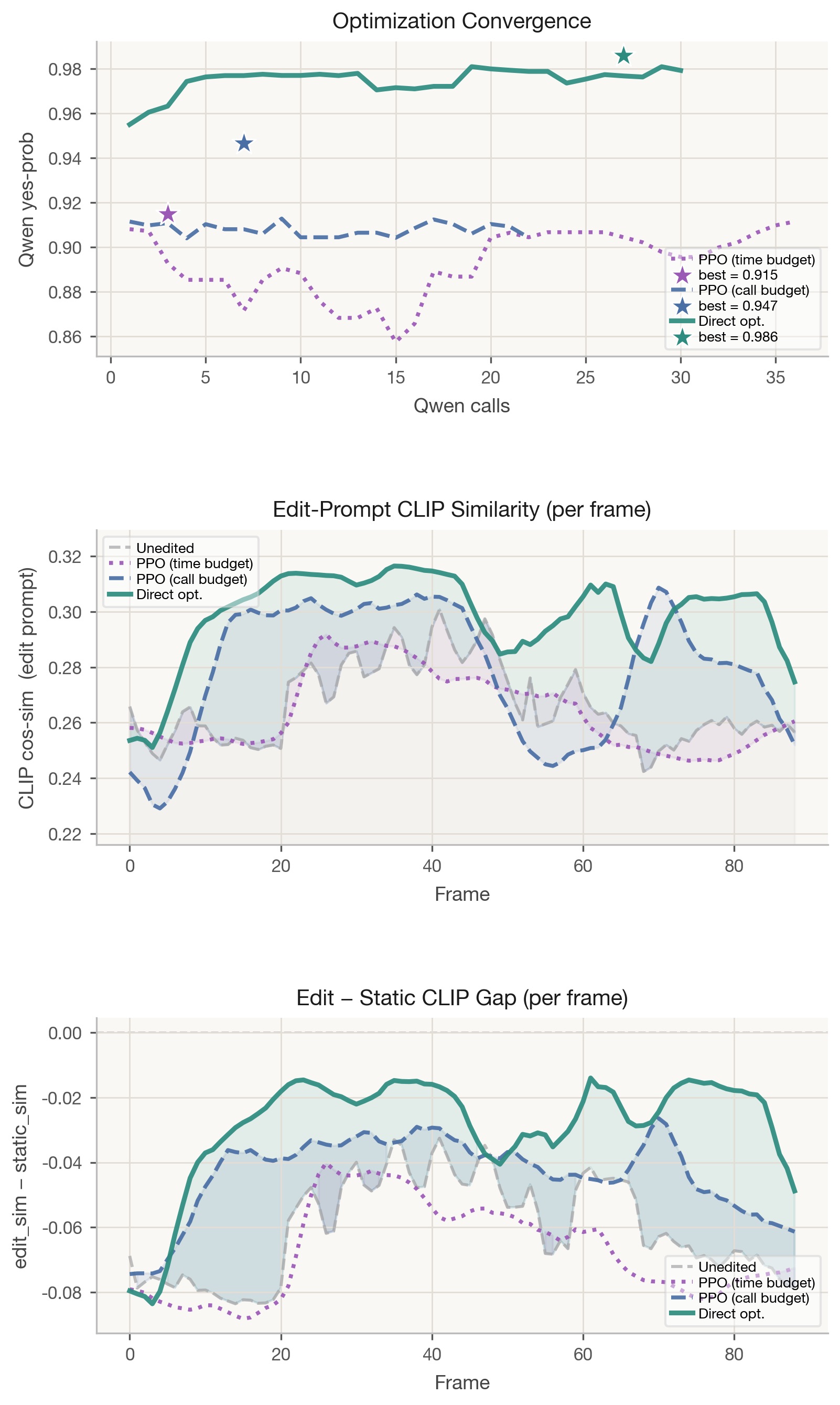}
    \vspace{-1.0em}
    \caption{
    \textbf{Direct latent optimization vs. PPO.}
    PPO is given the same variables, objective, renderer, and regularizers as our method, but treats editing as black-box reward maximization. Under matched Qwen-call and wall-clock budgets, direct gradient-based tuning discovers stronger motion edits more efficiently.
    }
    \label{fig:supp_ppo_ablation}
    \vspace{-1.0em}
\end{figure*}

We evaluate PPO under two matched-budget settings: the same number of Qwen calls as our optimizer, and the same optimization-loop wall-clock time. Fig.~\ref{fig:supp_ppo_ablation} compares direct optimization against PPO. The inset below shows a representative PPO output under the matched budget.

\vspace{0.35em}
\begin{center}
    \includegraphics[width=0.95\linewidth]{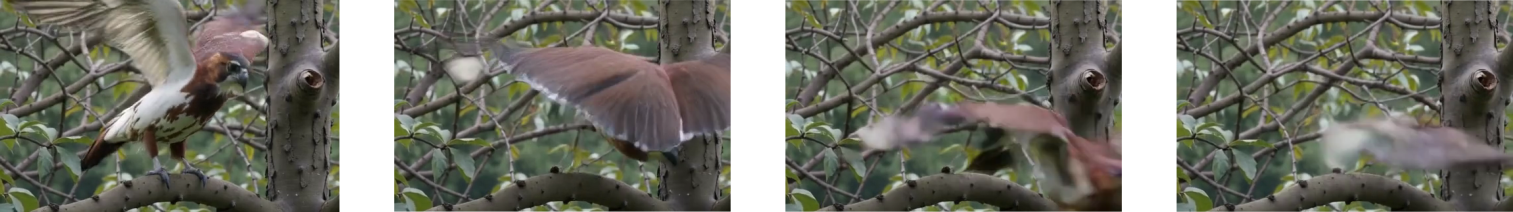}
\end{center}
\vspace{-0.35em}

As suggested by the inset, PPO can introduce artifacts and partial visual collapse under the same budget. This is expected: although PPO is scored with the same VLM objective and regularizers, it only receives a final scalar reward after each full-video rollout. It cannot exploit the directional gradients from the differentiable VLM loss, latent regularization, LPIPS preservation, or temporal preservation. Direct optimization is therefore more sample-efficient, using dense gradients through Qwen and the differentiable Retake path, while PPO performs costly black-box search in a high-dimensional continuous latent space. PPO remains reasonable when gradients are unavailable, but is less effective when our differentiable path is available.


\subsection{Latent Transferability}
\label{sec:supp_latent_transferability}

To analyze whether the optimized latents capture reusable motion directions, we evaluate transfer beyond the source clip. We first optimize the audio latent and residual text-conditioning vector for a source video and motion prompt, obtaining $\{\mathbf{\alpha}^{*}, \Delta\mathbf{v}^{*}\}$, then reuse them on a different target video with a related edit prompt without additional test-time tuning.

Fig.~\ref{fig:supp_transferability_quantitative} provides a CLIP-based analysis of this behavior. For each frame, we measure similarity to the target edit prompt and to a static prompt describing the source video. Successful transfer should increase edit-prompt alignment while reducing source-prompt alignment; the right plots show this difference directly, where positive values indicate stronger alignment with the target edit.

Across both examples, transferred latents yield a positive edit--source similarity gap over most frames, while the prompt-only baseline remains weaker or negative. This suggests that the optimized latents do not simply memorize one source clip, but encode motion-relevant directions that transfer to related subjects and scenes.

\subsection{Explanation of Modality Ablations}
\label{sec:supp_modality_explanation}

Tab.~1 of main paper evaluates the role of each tuning conditioning branch using GPT-5.5-Thinking as a VLM evaluator over 10 motion-centric scenarios. For each scenario, the evaluator receives the source video, edit prompt, and four outputs: text-only tuning, audio-only tuning, joint audio--text tuning without regularization, and the full joint model with regularization. Each result is scored on a 1--10 scale along multiple axes. \textit{Alignment} measures whether the generated video semantically matches the requested edit prompt and edits the correct subject or object. \textit{Motion strength} measures how clearly and completely the requested temporal change is expressed. \textit{Naturalness} measures temporal smoothness, physical plausibility, and whether the motion unfolds gradually rather than through abrupt jumps or sudden insertions. \textit{Preservation} measures how well the source identity, layout, background, and unrelated content are maintained. \textit{Visual quality} penalizes blur, flicker, distortions, and geometry artifacts, while \textit{edit locality} measures whether the change remains confined to the intended region without unwanted global drift.

The overall score used for this ablation is a weighted average that emphasizes prompt-aligned motion while still penalizing visual degradation and source drift. This modality ablation uses a more fine-grained scoring rubric than the baseline comparison prompt in Fig.~\ref{fig:supp_vlm_eval_prompt}, separating motion strength, naturalness, and edit locality:
\begin{equation}
\begin{aligned}
S_{\mathrm{overall}} =\;&
0.25 S_{\mathrm{align}}
+ 0.25 S_{\mathrm{motion}}
+ 0.20 S_{\mathrm{natural}} \\
&+ 0.15 S_{\mathrm{preserve}}
+ 0.10 S_{\mathrm{visual}}
+ 0.05 S_{\mathrm{local}} .
\end{aligned}
\end{equation}
This weighting reflects the goal of motion-centric editing: a successful result should not only match the text prompt, but also realize the requested action as a coherent temporal process while preserving the source video.

Fig.~\ref{fig:supp_text_vs_joint} provides a per-scenario breakdown comparing joint audio--text tuning against the text-only baseline. The left plot shows that joint tuning improves or matches text-only tuning in most scenarios, with the largest gains appearing in cases where motion quality and temporal stability are important, such as shouting, blooming, water interaction, and turtle-neck extension. The unregularized joint variant can also improve prompt alignment, but is less stable across scenarios, illustrating why regularization is needed when both latent spaces are tuned together. The right plot shows that joint tuning yields only a modest gain in semantic alignment and does not necessarily increase raw motion strength, but it substantially improves naturalness, source preservation, visual quality, and edit locality. This supports our interpretation that the audio latent is most useful as a temporally structured control signal that stabilizes and naturalizes the edit, rather than simply amplifying motion.

\section{Additional Qualitative Results}
\label{sec:supp_extended_qual}

\subsection{Comparison with All Baselines}
\label{sec:supp_full_baseline_comparison}

Fig.~\ref{fig:supp_qualitative_all} shows additional comparisons against all baselines. These examples further illustrate the main trend observed in the paper: prompt-only and general-purpose editors can either miss the requested temporal change or introduce unwanted scene changes, while \paper{} more consistently performs the target motion and preserves the source content.

\subsection{Comparison with Prompt-Only LTX Retake}
\label{sec:supp_ours_vs_ltx}

\begin{figure*}[t]
    \centering

    \includegraphics[width=\textwidth]{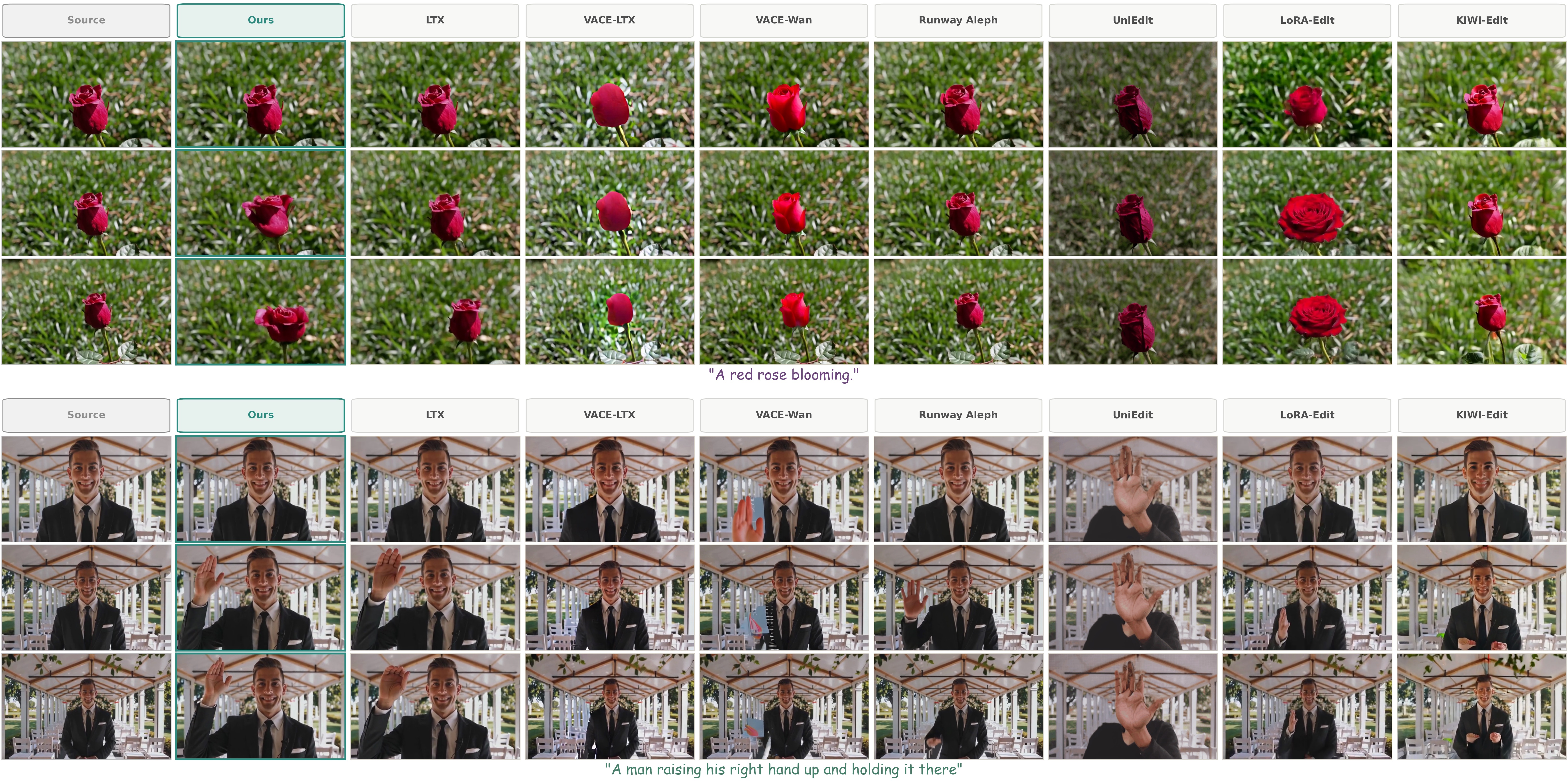}
    \vspace{-1.2em}

    \caption{
    \textbf{Additional qualitative comparisons.}
    Two additional motion-centric editing scenarios comparing our method with all baselines. Our method more reliably performs the requested motion without abrupt changes or artifacts while preserving the source scene.
    }
    \label{fig:supp_qualitative_all}

    \vspace{0.5em}

    \includegraphics[width=\textwidth]{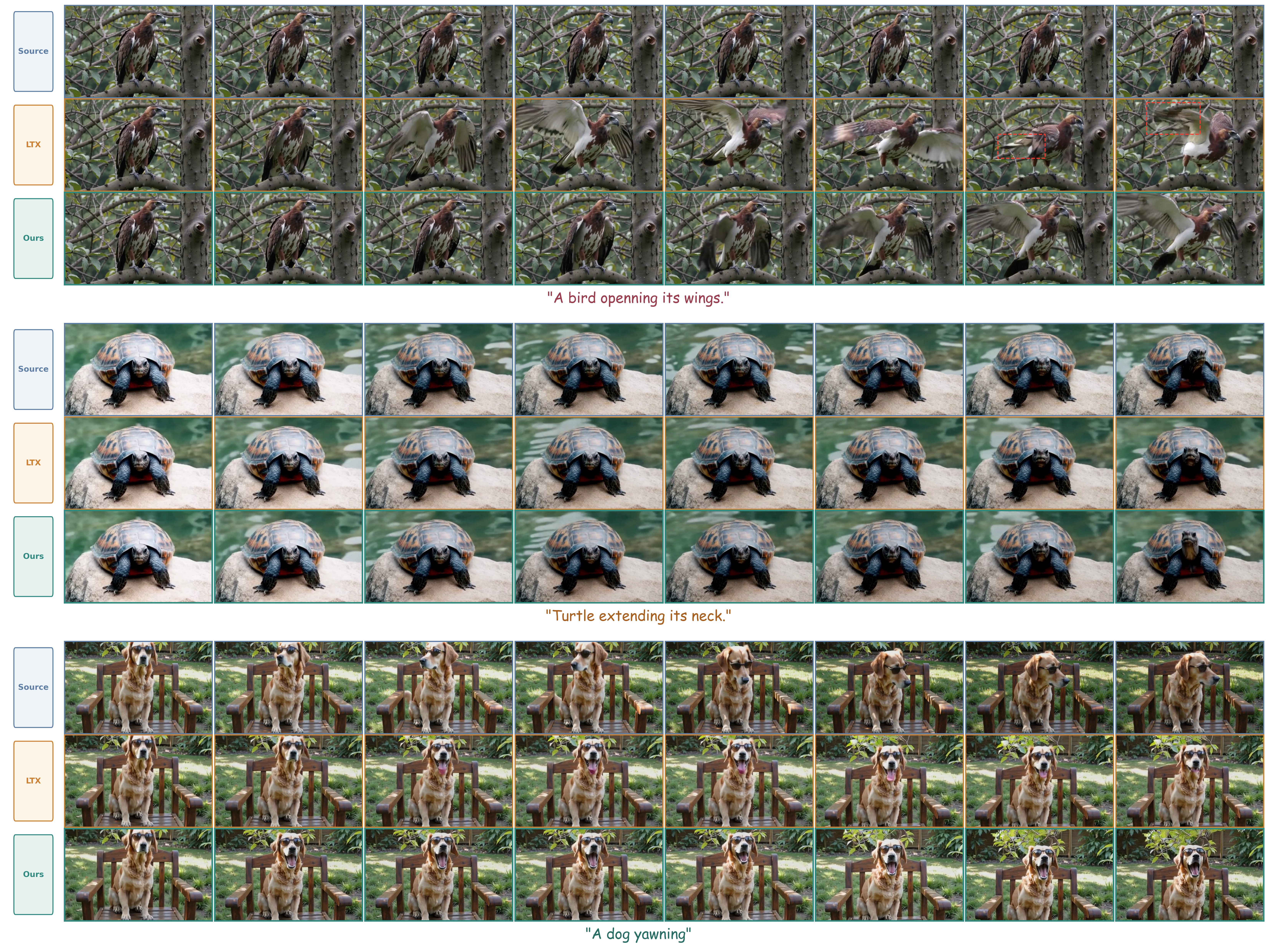}
    \vspace{-1.0em}

    \caption{
    \textbf{Additional comparison with LTX Retake.}
    Three additional scenarios comparing our method with prompt-only LTX Retake. Tuning internal conditioning latents improves motion realization while retaining source content.
    }
    \label{fig:supp_qualitative_ours_vs_ltx}

    \vspace{-1.0em}
\end{figure*}

Fig.~\ref{fig:supp_qualitative_ours_vs_ltx} compares our method directly with prompt-only LTX Retake on additional examples. The comparison isolates the effect of our test-time conditioning optimization: both methods use the same frozen backbone, but optimizing the internal audio and residual text latents improves motion realization.


\section{Benchmark and Evaluation Details}
\label{sec:supp_benchmark}

\subsection{Source Video Collection}
\label{sec:supp_source_videos}

We evaluate \paper{} on 25 motion-centric editing tasks: 20 AI-generated source videos and 5 real videos. The generated clips provide controlled scenes with clear subjects and unambiguous prompts, while the real videos add diversity in appearance, camera motion, background complexity, and natural image statistics.

Among the real videos, one clip is taken from the \emph{Neural 3D Video Dataset} introduced by Li et al.~\cite{li2022neural}, using one viewpoint from the commonly used ``cook spinach'' scene. The remaining four real clips are downloaded from Pexels\footnote{\url{https://www.pexels.com/}}, which provides freely downloadable stock videos. These real examples are used only as source videos for evaluating motion-editing behavior.

We run \paper{} on all 25 tasks. Since running all external baselines is computationally expensive, we select a representative subset of 12 tasks for the full comparison against all baselines. This subset is chosen to cover variation in object category, motion type, scene structure, source realism, and edit difficulty. It includes both generated and real inputs, and spans articulated motion, object-state changes, object--scene interactions, and localized temporal edits.

\begin{figure}[t]
    \centering
    \includegraphics[width=\linewidth]{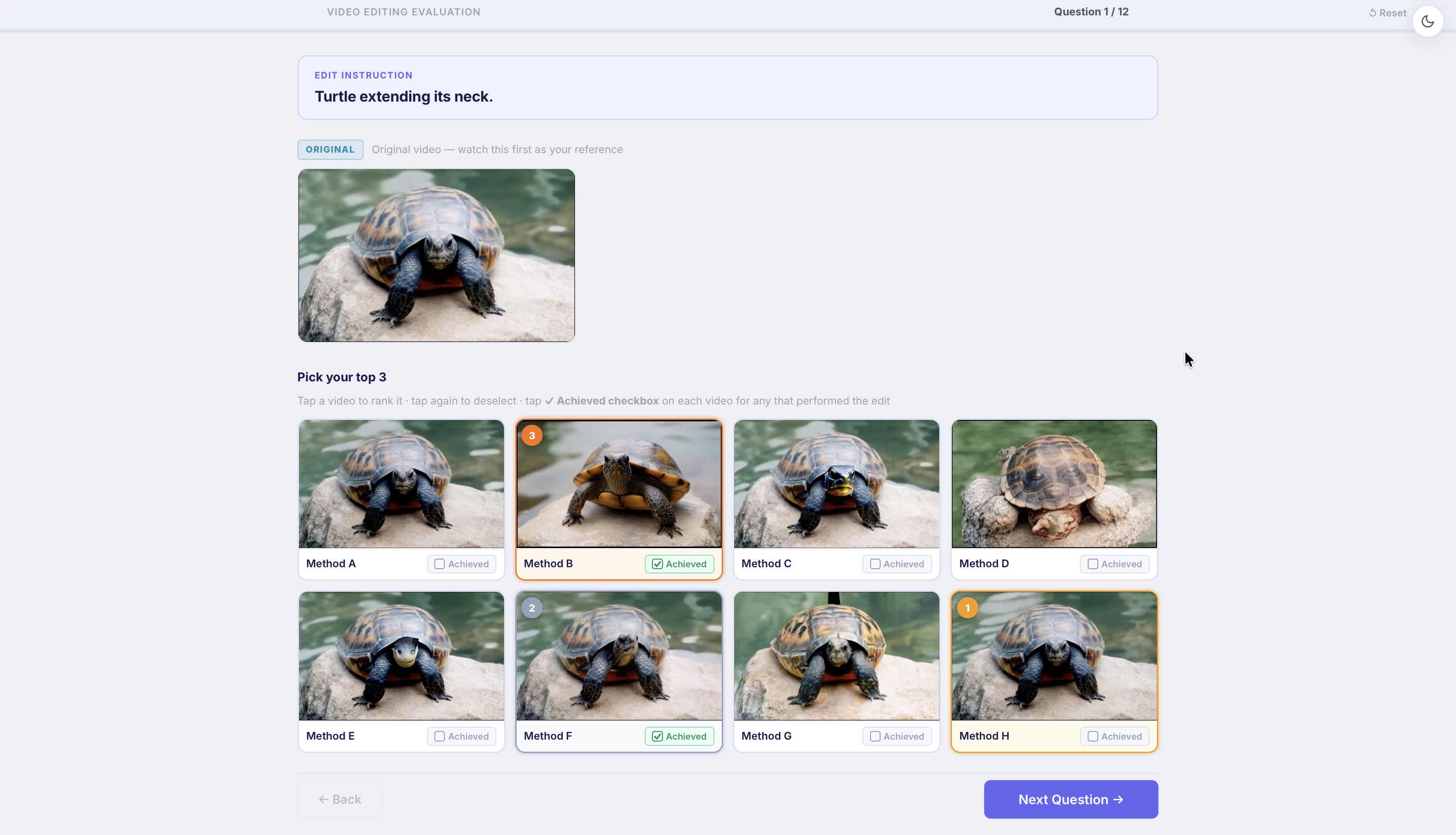}
    \vspace{-1.2em}
    \caption{
    \textbf{Human evaluation interface.}
    Participants view the edit instruction and source video, then rank their top three anonymized candidate edits and mark any videos that clearly achieve the requested motion using the \emph{Achieved} checkbox.
    }
    \label{fig:supp_human_survey_interface}
    \vspace{-1.2em}
\end{figure}

\subsection{Human Survey Interface}
\label{sec:supp_survey}

Fig.~\ref{fig:supp_human_survey_interface} shows the human-evaluation interface. For each scenario, participants see the edit instruction, source video, and 8 anonymized candidate edits corresponding to \paper{} and the 7 baselines. To reduce bias, scenario and candidate order are randomized per participant, and methods are shown only as Method A--H.

Participants select their top three edited videos and independently mark an \emph{Achieved} checkbox for candidates that clearly perform the requested motion or state change. After removing unreliable submissions with unrealistically short completion times, we compute Win Rate, Top-3 Rate, AVG Score using 3/2/1 points, and Task Achieved from the remaining responses.

\subsection{VLM Evaluation Prompt}
\label{sec:supp_vlm_eval_prompt}

For reproducibility, Fig.~\ref{fig:supp_vlm_eval_prompt} gives the exact GPT-based VLM evaluation prompt. For each scenario, \texttt{[Edit prompt]} and \texttt{[Video file names]} are replaced with the corresponding instruction and anonymized candidate names. The evaluator receives the source video and all candidate edits, then returns structured JSON scores and rankings.

\begin{figure*}[p]
\centering
\begin{tcolorbox}[
    width=0.82\textwidth,
    colback=gray!3,
    colframe=gray!35,
    title=\textbf{VLM evaluation prompt},
    fonttitle=\bfseries,
    boxrule=0.4pt,
    arc=2pt,
    left=5pt,
    right=5pt,
    top=4pt,
    bottom=4pt,
    enhanced
]
\begin{PromptVerbatim}
You are an expert evaluator for text-guided video editing.

You will receive:
1. One source video named `input_video`.
2. One edit prompt.
3. Multiple edited candidate videos. Each candidate has a unique video name.

Your task is to evaluate each candidate edited video against the source video and the edit prompt, then compare the candidates against each other.

Edit prompt:
`[Edit prompt]`

Candidate video names:
`[Video file names]`

UniEdit's output is a gif file, on the left is input video, and on the right is the edited result by UniEdit. Evaluate the right part, ignoring the left part.

Evaluate every candidate using these criteria:

A. Task achieved: `true` or `false`.
`true` only if the requested action, motion, or state change is clearly visible in the edited video.

B. Edit alignment: integer from 0 to 10.
How well does the edited video match the requested edit prompt?

C. Motion quality: integer from 0 to 10.
How complete, clear, temporally coherent, and natural is the induced motion or state change?

D. Source preservation: integer from 0 to 10.
How well are the original subject identity, scene layout, background, and unrelated content preserved from the source video?

E. Visual quality: integer from 0 to 10.
How free is the video from artifacts, flicker, distortions, identity collapse, and unnatural texture changes?

F. Overall score: integer or decimal from 0 to 10.
Compute this as:

`0.35 * edit_alignment + 0.25 * motion_quality + 0.20 * source_preservation + 0.20 * visual_quality`

Evaluation rules:
- Use the full 0 to 10 scale.
- Be strict and objective.
- Score each candidate independently against the source video and edit prompt before ranking candidates.
- Do not give a high edit alignment score if the requested motion or state change is missing.
- Do not give a high motion quality score if the motion is incomplete, flickery, temporally inconsistent, or physically implausible.
- Do not give a high source preservation score if the subject identity, layout, background, camera framing, or unrelated content changes unnecessarily.
- Do not give a high visual quality score if the video has strong artifacts, distortions, flicker, identity collapse, or unnatural texture changes.
- If a candidate preserves the source by barely changing anything, reward preservation but penalize task achievement, edit alignment, and motion quality.
- If a candidate performs the motion but damages identity, scene structure, or visual quality, reward the visible edit but penalize preservation and quality.
- Prefer candidates that satisfy the edit prompt while preserving source content and maintaining temporal/visual quality.
- Check the start, middle, and end of the video. Some videos may show motion in the end, some may show them spread through the video, and some may show them only in the middle. Consider the entire video when scoring.
- Penalize magically added/inserted objects. If something needs to be added by motion, it should happen naturally through the video, not appear suddenly without natural motion. This magically added content should be penalized in edit alignment, motion quality, and visual quality.

Return only valid JSON. Use the candidate video names exactly as provided.

Required JSON schema:

{
  "edit_prompt": "{EDIT_PROMPT}",
  "candidate_scores": [
    {
      "video_name": "candidate_name",
      "task_achieved": true,
      "edit_alignment": 0,
      "motion_quality": 0,
      "source_preservation": 0,
      "visual_quality": 0,
      "overall_score": 0.0,
      "reason": "Two concise sentences explaining the judgment."
    }
  ],
  "comparative_ranking": [
    "best_candidate_name",
    "second_best_candidate_name"
  ],
  "pairwise_preferences": [
    {
      "candidate_a": "candidate_name_a",
      "candidate_b": "candidate_name_b",
      "winner": "candidate_name_a or candidate_name_b or Tie",
      "reason": "One concise sentence."
    }
  ],
  "summary": "Two concise sentences summarizing the main differences between candidates."
}
\end{PromptVerbatim}
\end{tcolorbox}
\vspace{-0.5em}
\caption{
\textbf{Prompt used for automatic VLM evaluation.}
The placeholders are replaced by the corresponding edit prompt and video file names for each evaluated scenario.
}
\label{fig:supp_vlm_eval_prompt}
\vspace{-1.0em}
\end{figure*}


\clearpage
\onecolumn

\begingroup
\begin{center}
\setlength{\tabcolsep}{12pt}
\renewcommand{\arraystretch}{0.98}

\begin{longtable}{c l c c c c}
\caption{
  \textbf{Per-scenario human evaluation results.}
  \textbf{Win\%}: ranked 1st by raters.
  \textbf{Top-3\%}: ranked in top 3.
  \textbf{Avg Score}: mean rank score (3/2/1/0).
  \textbf{Task Ach.}: task achievement mean rate.
  Best per column \textbf{bold}; second-best \underline{underlined}.
}
\label{tab:human_survey_per_scenario}\\

\toprule
\textbf{Scenario} & \textbf{Method}
& \textbf{Win\% $\uparrow$}
& \textbf{Top-3\% $\uparrow$}
& \textbf{Avg Score $\uparrow$}
& \textbf{Task Ach. $\uparrow$} \\
\midrule
\endfirsthead

\caption[]{\textbf{Per-scenario human evaluation results} continued.}\\
\toprule
\textbf{Scenario} & \textbf{Method}
& \textbf{Win\% $\uparrow$}
& \textbf{Top-3\% $\uparrow$}
& \textbf{Avg Score $\uparrow$}
& \textbf{Task Ach. $\uparrow$} \\
\midrule
\endhead

\midrule
\multicolumn{6}{r}{\textit{Continued on next page}}\\
\endfoot

\bottomrule
\endlastfoot

\multirow{8}{*}{\rotatebox[origin=c]{90}{\textit{Car lights}}}
& \cellcolor{green!8}\textbf{Ours} & \cellcolor{green!8}\textbf{0.86} & \cellcolor{green!8}\textbf{1.00} & \cellcolor{green!8}\textbf{2.81} & \cellcolor{green!8}\textbf{0.86} \\
& LoRA Edit & 0.00 & 0.14 & 0.14 & 0.14 \\
& VACE (WAN 1.3) & 0.00 & 0.00 & 0.00 & 0.00 \\
& VACE (LTX) & 0.00 & 0.00 & 0.00 & 0.00 \\
& LTX2 Baseline & 0.05 & \underline{1.00} & \underline{1.76} & \underline{0.50} \\
& Runway / Aleph & \underline{0.10} & 0.86 & 1.29 & 0.29 \\
& Kiwi Edit & 0.00 & 0.00 & 0.00 & 0.00 \\
& UniEdit & 0.00 & 0.00 & 0.00 & 0.00 \\

\midrule

\multirow{8}{*}{\rotatebox[origin=c]{90}{\textit{Rose blooming}}}
& \cellcolor{green!8}\textbf{Ours} & \cellcolor{green!8}\underline{0.29} & \cellcolor{green!8}\underline{0.95} & \cellcolor{green!8}\underline{2.19} & \cellcolor{green!8}\underline{0.57} \\
& LoRA Edit & \textbf{0.71} & \textbf{1.00} & \textbf{2.71} & \textbf{0.86} \\
& VACE (WAN 1.3) & 0.00 & 0.10 & 0.10 & 0.07 \\
& VACE (LTX) & 0.00 & 0.00 & 0.00 & 0.00 \\
& LTX2 Baseline & 0.00 & 0.33 & 0.33 & 0.00 \\
& Runway / Aleph & 0.00 & 0.38 & 0.43 & 0.00 \\
& Kiwi Edit & 0.00 & 0.14 & 0.14 & 0.00 \\
& UniEdit & 0.00 & 0.10 & 0.10 & 0.00 \\

\midrule

\multirow{8}{*}{\rotatebox[origin=c]{90}{\textit{Goldfish}}}
& \cellcolor{green!8}\textbf{Ours} & \cellcolor{green!8}\textbf{0.62} & \cellcolor{green!8}\textbf{1.00} & \cellcolor{green!8}\textbf{2.52} & \cellcolor{green!8}\textbf{0.79} \\
& LoRA Edit & \underline{0.33} & \underline{1.00} & \underline{2.24} & \underline{0.71} \\
& VACE (WAN 1.3) & 0.00 & 0.00 & 0.00 & 0.00 \\
& VACE (LTX) & 0.00 & 0.00 & 0.00 & 0.00 \\
& LTX2 Baseline & 0.05 & 0.90 & 1.14 & 0.43 \\
& Runway / Aleph & 0.00 & 0.05 & 0.05 & 0.00 \\
& Kiwi Edit & 0.00 & 0.00 & 0.00 & 0.00 \\
& UniEdit & 0.00 & 0.05 & 0.05 & 0.00 \\

\midrule

\multirow{8}{*}{\rotatebox[origin=c]{90}{\textit{Man climbs rock}}}
& \cellcolor{green!8}\textbf{Ours} & \cellcolor{green!8}\textbf{0.57} & \cellcolor{green!8}\textbf{1.00} & \cellcolor{green!8}\textbf{2.57} & \cellcolor{green!8}\textbf{0.82} \\
& LoRA Edit & 0.00 & 0.81 & 1.14 & 0.45 \\
& VACE (WAN 1.3) & 0.00 & 0.05 & 0.05 & 0.18 \\
& VACE (LTX) & 0.00 & 0.00 & 0.00 & 0.00 \\
& LTX2 Baseline & 0.10 & 0.24 & 0.43 & 0.18 \\
& Runway / Aleph & \underline{0.33} & \underline{0.90} & \underline{1.81} & \underline{0.82} \\
& Kiwi Edit & 0.00 & 0.00 & 0.00 & 0.00 \\
& UniEdit & 0.00 & 0.00 & 0.00 & 0.00 \\

\midrule

\multirow{8}{*}{\rotatebox[origin=c]{90}{\textit{Man pets dog}}}
& \cellcolor{green!8}\textbf{Ours} & \cellcolor{green!8}\textbf{1.00} & \cellcolor{green!8}\textbf{1.00} & \cellcolor{green!8}\textbf{3.00} & \cellcolor{green!8}\textbf{0.79} \\
& LoRA Edit & \underline{0.00} & 0.14 & 0.29 & 0.00 \\
& VACE (WAN 1.3) & 0.00 & 0.29 & 0.38 & 0.00 \\
& VACE (LTX) & 0.00 & 0.05 & 0.10 & \underline{0.07} \\
& LTX2 Baseline & 0.00 & 0.67 & 0.95 & 0.00 \\
& Runway / Aleph & 0.00 & \underline{0.71} & \underline{1.05} & 0.07 \\
& Kiwi Edit & 0.00 & 0.05 & 0.05 & 0.00 \\
& UniEdit & 0.00 & 0.10 & 0.19 & 0.07 \\

\midrule

\multirow{8}{*}{\rotatebox[origin=c]{90}{\textit{Man raises hand}}}
& \cellcolor{green!8}\textbf{Ours} & \cellcolor{green!8}\textbf{0.62} & \cellcolor{green!8}\textbf{0.95} & \cellcolor{green!8}\textbf{2.38} & \cellcolor{green!8}\textbf{0.86} \\
& LoRA Edit & 0.10 & 0.43 & 0.81 & 0.71 \\
& VACE (WAN 1.3) & 0.00 & 0.05 & 0.05 & 0.00 \\
& VACE (LTX) & 0.00 & 0.00 & 0.00 & 0.00 \\
& LTX2 Baseline & 0.00 & 0.62 & 0.95 & 0.57 \\
& Runway / Aleph & \underline{0.29} & \underline{0.95} & \underline{1.81} & \underline{0.79} \\
& Kiwi Edit & 0.00 & 0.00 & 0.00 & 0.00 \\
& UniEdit & 0.00 & 0.00 & 0.00 & 0.00 \\

\midrule

\multirow{8}{*}{\rotatebox[origin=c]{90}{\textit{Bird opens wing}}}
& \cellcolor{green!8}\textbf{Ours} & \cellcolor{green!8}0.10 & \cellcolor{green!8}\textbf{0.95} & \cellcolor{green!8}\underline{1.62} & \cellcolor{green!8}0.67 \\
& LoRA Edit & 0.00 & 0.00 & 0.00 & 0.00 \\
& VACE (WAN 1.3) & 0.00 & 0.00 & 0.00 & 0.00 \\
& VACE (LTX) & 0.00 & 0.00 & 0.00 & 0.00 \\
& LTX2 Baseline & \textbf{0.67} & \underline{0.95} & \textbf{2.52} & \textbf{1.00} \\
& Runway / Aleph & \underline{0.24} & 0.86 & 1.57 & \underline{0.75} \\
& Kiwi Edit & 0.00 & 0.24 & 0.29 & 0.50 \\
& UniEdit & 0.00 & 0.00 & 0.00 & 0.00 \\

\midrule

\multirow{8}{*}{\rotatebox[origin=c]{90}{\textit{Dog yawning}}}
& \cellcolor{green!8}\textbf{Ours} & \cellcolor{green!8}\textbf{0.62} & \cellcolor{green!8}\textbf{1.00} & \cellcolor{green!8}\textbf{2.48} & \cellcolor{green!8}\textbf{0.92} \\
& LoRA Edit & \underline{0.24} & \underline{0.95} & \underline{2.05} & \underline{0.92} \\
& VACE (WAN 1.3) & 0.00 & 0.05 & 0.05 & 0.00 \\
& VACE (LTX) & 0.00 & 0.00 & 0.00 & 0.00 \\
& LTX2 Baseline & 0.05 & 0.67 & 0.90 & 0.42 \\
& Runway / Aleph & 0.00 & 0.00 & 0.00 & 0.00 \\
& Kiwi Edit & 0.00 & 0.10 & 0.10 & 0.25 \\
& UniEdit & 0.10 & 0.24 & 0.43 & 0.50 \\

\midrule

\multirow{8}{*}{\rotatebox[origin=c]{90}{\textit{Monkey reaches fruit}}}
& \cellcolor{green!8}\textbf{Ours} & \cellcolor{green!8}\textbf{0.52} & \cellcolor{green!8}\underline{0.76} & \cellcolor{green!8}\textbf{1.90} & \cellcolor{green!8}\underline{0.50} \\
& LoRA Edit & 0.19 & 0.62 & 1.24 & 0.36 \\
& VACE (WAN 1.3) & 0.00 & 0.05 & 0.10 & 0.07 \\
& VACE (LTX) & 0.00 & 0.00 & 0.00 & 0.00 \\
& LTX2 Baseline & 0.05 & \textbf{0.81} & 1.19 & 0.50 \\
& Runway / Aleph & \underline{0.24} & 0.76 & \underline{1.57} & \textbf{0.64} \\
& Kiwi Edit & 0.00 & 0.00 & 0.00 & 0.00 \\
& UniEdit & 0.00 & 0.00 & 0.00 & 0.00 \\

\midrule

\multirow{8}{*}{\rotatebox[origin=c]{90}{\textit{Red car door}}}
& \cellcolor{green!8}\textbf{Ours} & \cellcolor{green!8}\underline{0.43} & \cellcolor{green!8}\underline{0.95} & \cellcolor{green!8}\underline{2.29} & \cellcolor{green!8}\textbf{0.85} \\
& LoRA Edit & \textbf{0.52} & \textbf{1.00} & \textbf{2.48} & \underline{0.85} \\
& VACE (WAN 1.3) & 0.00 & 0.05 & 0.05 & 0.00 \\
& VACE (LTX) & 0.00 & 0.00 & 0.00 & 0.00 \\
& LTX2 Baseline & 0.00 & 0.00 & 0.00 & 0.00 \\
& Runway / Aleph & 0.05 & 0.24 & 0.38 & 0.38 \\
& Kiwi Edit & 0.00 & 0.76 & 0.81 & 0.54 \\
& UniEdit & 0.00 & 0.00 & 0.00 & 0.00 \\

\midrule

\multirow{8}{*}{\rotatebox[origin=c]{90}{\textit{Robot wave}}}
& \cellcolor{green!8}\textbf{Ours} & \cellcolor{green!8}\textbf{0.67} & \cellcolor{green!8}\textbf{1.00} & \cellcolor{green!8}\textbf{2.57} & \cellcolor{green!8}\textbf{1.00} \\
& LoRA Edit & 0.00 & 0.00 & 0.00 & 0.00 \\
& VACE (WAN 1.3) & 0.00 & 0.24 & 0.24 & 0.58 \\
& VACE (LTX) & 0.00 & 0.00 & 0.00 & 0.00 \\
& LTX2 Baseline & \underline{0.29} & \underline{1.00} & \underline{2.29} & \underline{1.00} \\
& Runway / Aleph & 0.05 & 0.76 & 0.90 & 0.83 \\
& Kiwi Edit & 0.00 & 0.00 & 0.00 & 0.00 \\
& UniEdit & 0.00 & 0.00 & 0.00 & 0.17 \\

\midrule

\multirow{8}{*}{\rotatebox[origin=c]{90}{\textit{Turtle extends neck}}}
& \cellcolor{green!8}\textbf{Ours} & \cellcolor{green!8}\textbf{0.48} & \cellcolor{green!8}\underline{0.86} & \cellcolor{green!8}\underline{2.14} & \cellcolor{green!8}\textbf{0.93} \\
& LoRA Edit & \underline{0.48} & \textbf{1.00} & \textbf{2.29} & \underline{0.79} \\
& VACE (WAN 1.3) & 0.00 & 0.14 & 0.19 & 0.43 \\
& VACE (LTX) & 0.00 & 0.10 & 0.10 & 0.21 \\
& LTX2 Baseline & 0.00 & 0.38 & 0.43 & 0.21 \\
& Runway / Aleph & 0.05 & 0.52 & 0.86 & 0.57 \\
& Kiwi Edit & 0.00 & 0.00 & 0.00 & 0.07 \\
& UniEdit & 0.00 & 0.00 & 0.00 & 0.00 \\

\end{longtable}
\end{center}
\endgroup

\clearpage
\twocolumn

\end{document}